\documentclass{aa}   
\usepackage{graphicx}
\usepackage{txfonts}
\usepackage{epsfig}
\usepackage{longtable}
\usepackage{lscape}
\usepackage{amssymb} 
\usepackage{natbib} 
\usepackage[colorlinks=true,citecolor=blue]{hyperref} 
\usepackage{refcount}

\def\xmmj {\mbox{2XMM~J185114.3$-$000004}} 
\def \igr {\mbox{IGR~J17407$-$2808}} 
\def \nosrc {\mbox{IGR~J18175$-$2419}} 

\def \chandra {{\em Chandra}}
\def \inte {{\em INTEGRAL}}

\def \suzaku {{\em Suzaku}}
\def \sw {{\em Swift}}
\def \xmm {{\em XMM--Newton}}

\def \hcm {\hbox {\ifmmode $ atom cm$^{-2}\else atom cm$^{-2}$\fi}}
\def \arcmin {\hbox{$^\prime$}}
\def \arcsec {\hbox{$^{\prime\prime}$}}

\def \aip {AIP Conf. Ser.}
\def \ATel {ATel} 
\def \apj {ApJ}
\def \apjl {ApJL}
\def \apjs {ApJS}\def \aap {A\&A}

\def \asr {AdSpR}

\def \gcn {GCN Circ.} 

\def \mnras {MNRAS}

\def \ssr {SSRv}

\begin{document} 

\title{Searching for supergiant fast X-ray transients with \emph{Swift}  } 
\titlerunning{Searching for SFXTs}
\authorrunning{Romano et al.}
\author{P.~Romano\inst{1}     
        \and 
       E.~Bozzo\inst{2} 
         \and
       P.~Esposito\inst{3}  
            \and
       B.~Sbarufatti\inst{4,5} 
          \and 
       F.~Haberl\inst{6} 
            \and
       G.~Ponti\inst{6} 
            \and
       P.~D'Avanzo\inst{4} 
          \and 
       L.~Ducci\inst{7,2} 
          \and 
       A.~Segreto\inst{1}  
           \and 
        C.~Jin\inst{6} 
           \and 
        N.~Masetti\inst{8,9}
            \and
        M.~Del~Santo\inst{1}     
           \and 
        S.~Campana\inst{4} 
         \and 
        V.~Mangano\inst{5}
            }
  \institute{INAF, Istituto di Astrofisica Spaziale e Fisica Cosmica - Palermo,
              Via U.\ La Malfa 153, I-90146 Palermo, Italy \\
              \email{romano@ifc.inaf.it}
               \and 
         ISDC Data Center for Astrophysics, Universit\'e de Gen\`eve, 16 chemin d'\'Ecogia, 1290 Versoix, Switzerland
                 \and 
               Anton Pannekoek Institute, University of Amsterdam, Postbus 94249, NL-1090-GE Amsterdam, The Netherlands
                   \and 
         INAF, Osservatorio Astronomico di Brera,  Via E.\ Bianchi 46, I-23807 Merate, Italy 
            \and
         Department of Astronomy and Astrophysics, Pennsylvania State 
              University, University Park, PA 16802, USA
              \and
         Max-Planck-Institut f\"ur extraterrestrische Physik, Giessenbachstra{\ss}e, 85748 Garching, Germany
                 \and 
         Institut f\"ur Astronomie und Astrophysik, Eberhard Karls Universit\"at, 
            Sand 1, 72076 T\"ubingen, Germany
              \and
              INAF, Istituto di Astrofisica Spaziale e Fisica Cosmica - Bologna, 
              Via Gobetti, 101,  I-40129 Bologna,  Italy 
              \and 
              Departamento de Ciencias F\'isicas, Universidad Andr\'es Bello,
            Fern\'andez Concha 700, Las Condes, Santiago, Chile
            }
\date{Received 28 April, 2016; accepted 18 June, 2016}

\abstract{
Supergiant fast X--ray transients (SFXTs) are high mass X--ray binaries (HMXBs) 
hosting a neutron star and an OB supergiant companion. 
We examine the available \sw\ data, as well as other new or archival/serendipitous data, on three sources:
IGR~J17407$-$2808, 2XMM~J185114.3$-$000004, and IGR~J18175$-$2419, 
whose X--ray characteristics qualify them as candidate SFXT, in order to explore their  properties 
and test whether they are  consistent with an SFXT nature. 
As IGR~J17407$-$2808 and 2XMM~J185114.3$-$000004 triggered the Burst Alert Telescope on board \sw, 
the \sw\ data allow us to provide their first arcsecond localisations, leading to 
an unequivocal identification of the source CXOU~J174042.0$-$280724 as the soft X--ray 
counterpart of  IGR~J17407$-$2808, as well as their first broadband spectra, which can be fit 
with models generally describing accreting neutron stars in HMXBs. 
While still lacking optical spectroscopy to assess the spectral type of the companion, 
we propose 2XMM~J185114.3$-$000004 as a very strong SFXT candidate. 
The nature of  IGR~J17407$-$2808 remains, instead, more uncertain. Its broad band properties 
cannot exclude that the emission originates from either a HMXB (and in that case, a SFXT) or, more likely,  a low mass X--ray binary. 
Finally, based on the deep non-detection in our XRT monitoring campaign and a careful reanalysis of the original 
\inte\ data in which the discovery of the source was first reported, we show that IGR~J18175$-$2419 is likely a 
spurious detection. 
}


   \maketitle
%

	\section{Introduction\label{sfxt3cand:intro}}

 \setcounter{table}{0}   
\begin{table*}[th] 	
\tabcolsep 4pt   
\begin{center} 	
 \caption{Log of X--ray observations of 2XMM~J185114.3$-$000004 and spectral fit results. } 	
 \label{sfxt3cand:tab:2xmmlog} 	
\scriptsize
\begin{tabular}{c rllr ccccc }
 \hline 
 \hline 
 \noalign{\smallskip}  
Instrument               &  ObsID                 & Start time                     & End time   & Exposure  & $N_{\rm H}$                             & $\Gamma$  & $F_{\rm 0.5-10\,keV}^{a}$    & $F_{\rm 15-50\,keV}^{a}$ & $\chi^2/$dof \\    
                                 &                            &   (UT)                             &    (UT)        & (s)             &  ($\times10^{22}$ cm$^{-2}$)&                     &   (erg cm$^{-2}$ s$^{-1}$) &   (erg cm$^{-2}$ s$^{-1}$) &  \\              
 \noalign{\smallskip} 
 \hline 
 \noalign{\smallskip} 
\sw/BAT evt & 00524542000	&2012-06-17 15:43:04 	&2012-06-17 17:58:19	&	1201  & --                                             & $2.6_{-0.4}^{+0.4}$                & --       &  $(7.8\pm0.8)\times$10$^{-10}$ & 26.1/24\\  
\sw/XRT WT & 00524542000	&2012-06-17 15:49:58	&2012-06-17 17:14:58	&	17	  & $^{b}$                           & $^{b}$                            &    $(3.4\pm0.7)\times$10$^{-10}$    &  -- & --          \\  
\sw/XRT PC & 00524542000	&2012-06-17 15:50:11	&2012-06-17 17:56:46	&	4305  & $14_{-3}^{+4}$             & $0.77_{-0.44}^{+0.49}$                  &     $(1.1^{+0.5}_{-0.1})\times$10$^{-10}$   & --  & 45.4/42 \\     
\sw/XRT PC & 00524542001	&2012-06-17 18:51:27	&2012-06-18 01:39:58	&	5366  & $^{b}$                           & $^{b}$                            &    $(5.4\pm0.2)\times$10$^{-13}$   &   --  & --          \\  
\sw/XRT PC & 00524542002	&2012-06-18 19:06:54	&2012-06-18 19:33:56	&	1610 & $^{b}$                           & $^{b}$                            & $<1.1\times$10$^{-12}$    & --   & 3$\sigma$\,UL	\\  
\sw/XRT PC & 00524542003	&2012-06-18 22:09:31	&2012-06-19 01:53:44	&	3831 & $^{b}$                            & $^{b}$                            & $<1.3\times$10$^{-12}$    & --   & 3$\sigma$\,UL	\\  
\sw/XRT PC & 00524542004	&2012-06-19 17:20:22	&2012-06-19 17:36:57	&	985	 & $^{b}$                            & $^{b}$                            & $<1.6\times$10$^{-12}$    & --   & 3$\sigma$\,UL \\  
\sw/XRT PC & 00524542005	&2012-06-21 03:09:28	&2012-06-21 03:26:59	&	1020 & $^{b}$                            & $^{b}$                            & $<2.7\times$10$^{-12}$    & --   & 3$\sigma$\,UL	\\  
\sw/XRT PC & 00524542007	&2012-06-23 04:47:40	&2012-06-23 04:56:55	&	554	 & $^{b}$                            & $^{b}$                            & $<3.2\times$10$^{-12}$    & --   & 3$\sigma$\,UL \\   
\sw/XRT PC & 00524542008	&2012-06-24 03:48:15	&2012-06-24 23:09:56	&	1158 & $^{b}$                            & $^{b}$                            & $<1.8\times$10$^{-12}$    & --   & 3$\sigma$\,UL	\\  
\sw/XRT PC & 00524542009	&2012-06-25 08:05:54	&2012-06-25 22:38:56	&	1131 & $^{b}$                            & $^{b}$                            &       $(2.6\pm0.8)\times$10$^{-12}$ & --   & --	                     \\  
\sw/XRT PC & 00524542010	 &2012-06-26 00:10:27	&2012-06-26 00:25:55	&	923	 & $^{b}$                            & $^{b}$                            & $<2.5\times$10$^{-12}$    & --   & 3$\sigma$\,UL\\  
\sw/XRT PC & 00524542011	&2012-06-28 10:00:53	&2012-06-28 10:17:56	&	1000 & $^{b}$                            & $^{b}$                            & $<2.7\times$10$^{-12}$     & --   & 3$\sigma$\,UL \\  
\sw/XRT PC & 00524542012	&2012-06-29 06:55:10	&2012-06-29 07:11:56	&	995	 & $^{b}$                            & $^{b}$                            & $<1.5<\times$10$^{-12}$   & --   & 3$\sigma$\,UL \\    
\sw/XRT PC & 00524542013	&2012-06-30 02:22:00	&2012-06-30 02:38:55	&	1003 & $^{b}$                            & $^{b}$                            & $<1.8<\times$10$^{-12}$   & --   & 3$\sigma$\,UL	\\    
\sw/XRT PC & 00524542014	&2012-07-01 03:55:57	&2012-07-01 03:59:55	&	228	 & $^{b}$                            & $^{b}$                            & $<6.7<\times$10$^{-12}$   & --   & 3$\sigma$\,UL \\    
\sw/XRT PC & 00524542015	&2012-07-02 03:48:53	&2012-07-02 04:01:56	&	757	 & $^{b}$                            & $^{b}$                            & $<2.4<\times$10$^{-12}$   & --   & 3$\sigma$\,UL \\    
\sw/XRT PC & 00524542016	&2012-07-03 01:04:48	&2012-07-03 02:26:54	&	1005 & $^{b}$                            & $^{b}$                            & $<1.8<\times$10$^{-12}$   & --   & 3$\sigma$\,UL	\\    
\sw/XRT PC & 00524542017	&2012-07-04 21:46:20	&2012-07-04 22:01:55	&	923	 & $^{b}$                            & $^{b}$                            & $<1.9<\times$10$^{-12}$   & --   & 3$\sigma$\,UL \\  
\noalign{\smallskip} 
\sw/XRT PC & 00032512001	&2012-08-08 01:19:37	&2012-08-08 19:08:56	&	2745 & $^{b}$                            & $^{b}$                            &        $(5.1\pm0.7)\times$10$^{-12}$  & --   & --	\\  
\sw/XRT PC & 00032512002	&2012-08-11 10:43:32	&2012-08-11 12:31:55	&	1464 & $^{b}$                            & $^{b}$                            & $<1.9\times$10$^{-12}$      & --   & 3$\sigma$\,UL	\\   
\sw/XRT PC & 00032512004  &2012-08-19 03:11:15	&2012-08-19 06:31:55	&	3879 & $^{b}$                            & $^{b}$                            & $<1.7\times$10$^{-12}$      & --   & 3$\sigma$\,UL	\\   
\sw/XRT PC & 00032512005	&2012-08-31 15:02:03	&2012-08-31 18:37:56	&	4719 & $^{b}$                            & $^{b}$                            &        $(3.4\pm0.4)\times$10$^{-12}$  & --   & --	\\  
 \noalign{\smallskip} 
\sw/XRT PC$^{c}$ & 00044565001	&2012-11-20 20:52:24	&2012-11-20 21:00:56 &506	& $^{b}$                            & $^{b}$                             & $<1.8\times$10$^{-12}$      & --   & 3$\sigma$\,UL\\  
\noalign{\smallskip} 
{\it XMM} Epic-pn$^{c}$     & 0017740401 &2003-10-05 00:53:54  &2003-10-05 08:57:14  &  27670 & $11\pm5$           & $1.6\pm0.9$         &$(3.3^{+0.7}_{-0.4})\times$10$^{-13}$   & --& 45.2/42$^{e}$ \\  
{\it XMM} Epic-MOS$^{c}$  & 0017740401 &2003-10-05 00:53:54  &2003-10-05 08:57:14  &  27670 & 13$^{+6}_{-4}$  & 2.0$^{+1.1}_{-0.9}$ &$(5.0^{+1.5}_{-1.0})\times$10$^{-13}$   & --& 15.2/19$^{e}$ \\      
  
{\it XMM} Epic-pn$^{c}$     & 0017740501 &2003-10-20 23:36:20  &2003-10-21 07:56:20  &  28667 & $6^{+12}_{-5}$   & $1.0^{+1.7}_{-1.3}$ &$(1.5^{+0.8}_{-0.4})\times$10$^{-13}$   & --& 9.6/14$^{e}$ \\    
{\it XMM} Epic-MOS$^{c}$  & 0017740501 &2003-10-20 23:36:20  &2003-10-21 07:56:20  &  28667 & 10$^{+14}_{-6}$ & 1.2$^{+2.0}_{-1.3}$ &$(1.5^{+1.5}_{-0.8})\times$10$^{-13}$  & --& 8.4/7$^{e}$ \\ 
{\it XMM} Epic-pn$^{c}$     & 0671510101 &2012-03-18 09:44:36  &2012-03-19 02:48:13  &  60084 & $^{d}$                         & $^{d}$                    & $<4\times$10$^{-15}$    & --& 3$\sigma$\,UL \\   
  \noalign{\smallskip}
  \hline
  \end{tabular}
  \end{center}
  \tablefoot{The adopted spectral model for the fits is an absorbed power law. 
$^a$:  Observed.    $^b$:  Adopting the same spectral parameters as in observation 0017740401 (pn).   $^c$:  Serendipitous observations. 
  $^d$:  Adopting the same spectral parameters as in observation 0017740501 (pn).    $^e$:  Cash statistics has been used during the fit.  }
  \end{table*}

Supergiant fast X--ray transients (SFXTs) are high mass X--ray binaries (HMXBs) 
hosting most likely a neutron star (NS) and 
an OB supergiant companion \citep{Sguera2005,Negueruela2006}. 
Differently from normal supergiant HMXBs, which display a fairly constant average 
luminosity with typical variations by a factor of 10--50 on time scales of a few hundred 
to thousands of seconds, SFXTs are characterised by hard X--ray flares reaching, for a few hours, 
10$^{36}$--10$^{37}$~erg~s$^{-1}$ \citep[see][for a catalog of hard X--ray flares]{Romano2014:sfxts_catI}. 
SFXTs have also been found to be significantly subluminous with respect to classical 
supergiant HMXBs like Vela X-1 \citep[][]{Lutovinov2013:HMXBpop,Bozzo2015:underluminous}, 
and show a soft X--ray dynamical range  up to 6 orders of magnitude 
\citep[][]{Sguera2005,Romano2015:17544sb}, as their luminosities can be as low as 
$\sim 10^{32}$~erg~s$^{-1}$ during quiescence 
\citep[e.g.][]{zand2005,Bozzo2010:quiesc1739n08408}. 
The origin of this different behaviour is still unknown 
\citep[see, e.g.][]{Bozzo2013:COSPAR_sfxt,Bozzo2015:underluminous} 
and thus the different models proposed to explain the behaviour for these sources 
are still debated. The models include a combination of 
more pronounced dense inhomogeneities (clumps) in the winds 
of the SFXT supergiant companions compared to those of classical systems 
\citep[][]{zand2005,Walter2007,Negueruela2008},
magnetic/cen\-trif\-u\-gal gates generated by the slower rotational 
velocities and higher magnetic fields of the NSs hosted in SFXTs 
\citep{Grebenev2007,Bozzo2008,Bozzo2016},  
or a subsonic settling accretion regime combined with magnetic reconnections between 
the NS and the supergiant stellar field transported by its wind 
\citep[][and references therein]{Shakura2014:bright_flares}. 

Most SFXTs were first discovered, or classified as such, 
based on their hard X--ray properties \citep[e.g.][]{Sguera2005,Negueruela2006,Walter2015:HMXBs_IGR} 
as observed by \inte/IBIS \citep[][]{Ubertini2003}. 
Subsequently, their long term behaviour has also been extensively investigated \citep[see][]{Romano2014:sfxts_catI}  
with other coded-mask large field-of-view instruments, such as the 
\sw/Burst Alert Telescope \citep[BAT,][]{Barthelmy2005:BAT}. 
These hard X--ray monitors, however, share similiar sensitivity limits, 
that can enable them to catch only the brightest portion of any transient event. 
Due to their transient nature, SFXTs are indeed particularly difficult to find unless
they experience frequent and relatively bright flares. 
A legitimate question is, therefore, whether we have already discovered the majority of Galactic SFXTs. 

In a recent work, \citet[][]{Ducci2014:sfxtN} have addressed the issue of how common the SFXT
phenomenon is, and concluded that, since detection of the entire population is hindered 
by the SFXT peculiar transient properties, a fraction of the population is probably 
yet to be identified, and that we are likely missing SFXTs with low outburst rates or large distances. 
\citet[][]{Ducci2014:sfxtN}, have considered two datasets, 
the 100 month \sw/BAT catalogue \citep[][]{Romano2014:sfxts_catI} 
and the first nine years of \inte/ISGRI data \citep[][]{Paizis2014}, 
applied two distinct statistical approaches to derive the expected number of SFXTs emitting bright flares in the Milky Way,
$N\approx 37{+53\atop-22}$. This value not only agrees with the expected number of 
HMXBs in the Galaxy derived from high-mass binary evolution studies 
\citep[][and references therein]{vandenHeuvel2012,Dalton1995}, but also 
suggests that SFXTs constitute a sizeable fraction of X--ray binaries with supergiant companions. 
The SFXT class currently includes a mere dozen confirmed individuals, that is, 
X--ray binaries for which optical/IR spectroscopy has firmly established the presence of 
a supergiant primary of O or B spectral type \citep[e.g.][]{Romano2014:sfxts_catI,Romano2015:swift10}. 
About as many candidate SFXTs are known, for which no optical spectroscopy has been obtained until now, 
but which have a reported history of bright, large dynamic range hard X--ray flaring. 
Since we expect a larger number of SFXTs in the Galaxy, 
it is worthwhile to increase the sample of these sources 
through new and archival multifrequency data studies of
SFXT candidates and other promising unclassified transients.
A larger sample of SFXTs would ultimately allow us to gain more information
to understand the accretion mechanisms responsible for their enigmatic behaviour.

The \emph{Swift}  SFXT Project \citep[][]{Romano2015:swift10} 
has been investigating the X--ray properties of the SFXTs since 2007, exploiting the unique 
observing capabilities of \sw\ \citep[][]{Gehrels2004}. 
In particular, we have been performing follow-up observations of several tens of SFXT 
outbursts caught by the BAT 
\citep[][and references therein]{Romano2011:sfxts_paperVII,Farinelli2012:sfxts_paperVIII,Romano2015:17544sb},  
with the X--ray Telescope \citep[XRT,][]{Burrows2005:XRT} 
and the UV/Optical Telescope \citep[UVOT,][]{Roming2005:UVOT}, 
and carried out long term monitorings of virtually all SFXTs, 
including also five classical systems for comparison purposes. 
\sw\ broad-band data of bright flares are particularly 
helpful in order to increase the SFXT sample, since they allow us to make a solid 
connection between the hard X--ray transient 
and its soft X--ray counterpart. 
As the XRT positional accuracy is as good as few arcseconds, such an association allows us 
to identify in most cases the optical/IR source associated with the X--ray transient and, 
subsequently, schedule dedicated optical spectroscopic campaigns to unveil its nature.

\setcounter{table}{1}  
 \begin{table*}
\tabcolsep 4pt   
\begin{center} 	
 \caption{Log of X--ray observations of IGR~J17407$-$2808 and spectral fit results. 
 \label{sfxt3cand:tab:17407log} }	
 \scriptsize
\begin{tabular}{c rllr ccccc }
 \hline 
 \hline 
 \noalign{\smallskip}  
Instrument               &  ObsID                 & Start time                     & End time   & Exposure & $N_{\rm H}$                           & $\Gamma$ & $F_{\rm 0.5-10\,keV}^{a}$   & $F_{\rm 15-50\,keV}^{a}$           & $\chi^2/$dof \\ 
                                 &                            &   (UT)                             &    (UT)       & (s)            &  ($\times10^{22}$ cm$^{-2}$) &                     & (erg cm$^{-2}$ s$^{-1}$) &  (erg cm$^{-2}$ s$^{-1}$)  & \\             
 \noalign{\smallskip} 
 \hline 
 \noalign{\smallskip} 
\sw/BAT evt &00505516000               &       2011-10-15 01:13:06   &       2011-10-15 01:28:11   &  905   & --                                    & $1.3_{-0.5}^{+0.5}$     & --                     & $ (4.6\pm0.9)\times$10$^{-10}$  & 49.6/24     \\
\sw/BAT evt (peak1) &00505516000  &       2011-10-15 01:12:26   &       2011-10-15  01:14:06  &  100   & --                                    & $2.1_{-0.5}^{+0.6}$     & --                     & $ (2.1\pm0.2)\times$10$^{-9}$   & 27.8/23     \\
\sw/BAT evt (peak2) &00505516000  &       2011-10-15 01:27:04   &       2011-10-15  01:28:49  &  105  & --                                     & $1.9_{-0.4}^{+0.4}$     & --                     & $ (2.3\pm0.3)\times$10$^{-9}$   & 32.8/24     \\
\sw/XRT PC  &00505516000               &       2011-10-15 01:15:05   &       2011-10-15 03:03:32   &  772   &  $0.84_{-0.84}^{+2.00}$$^{b}$  & $-0.45_{-0.71}^{+0.86}$  & $ (6.8\pm0.9)\times$10$^{-11}$  & --                   & 68.9/68$^{c}$    \\
 \noalign{\smallskip} 
\sw/XRT PC &00036122001   &       2011-10-15 19:16:28     &       2011-10-15 20:27:37     &       983    & $^{d}$                          & $^{d}$                          & $<2.1\times$10$^{-12}$    & --   & 3$\sigma$\,UL   \\
\sw/XRT PC &00036122002   &       2011-10-17 01:17:25     &       2011-10-17 01:34:01     &       988    & $^{d}$                           & $^{d}$                           & $<2.1\times$10$^{-12}$    & --   & 3$\sigma$\,UL  \\
\sw/XRT PC &00036122003   &       2011-10-17 21:57:29     &       2011-10-17 22:14:05     &       980    & $^{d}$                           & $^{d}$                          & $<2.7\times$10$^{-12}$    & --   & 3$\sigma$\,UL  \\
\sw/XRT PC &00036122004   &       2011-10-18 12:36:46     &       2011-10-18 12:53:57     &       1018  & $^{d}$                          & $^{d}$                          & $<2.9\times$10$^{-12}$    & --   & 3$\sigma$\,UL  \\
\sw/XRT PC &00036122006   &       2011-10-20 12:47:59     &       2011-10-20 12:48:27     &       28      & $^{d}$                           & $^{d}$                           & $<4.0\times$10$^{-11}$    & --   & 3$\sigma$\,UL \\
\sw/XRT PC &00036122007   &       2011-10-21 14:26:20     &       2011-10-21 14:41:56     &       933    & $^{d}$                           & $^{d}$                         & $<1.6\times$10$^{-12}$    & --   & 3$\sigma$\,UL \\
 \noalign{\smallskip} 
\sw/XRT PC$^{e}$ &00032370001  &2012-06-30 03:34:12  &2012-06-30 03:43:56 &      584    & $^{d}$                         & $^{d}$                          & $<4.5\times$10$^{-12}$    & --   & 3$\sigma$\,UL \\
\sw/XRT PC$^{e}$ &00032370003  &2012-07-12 11:09:06  &2012-07-12 11:17:55 &      522    & $^{d}$                          & $^{d}$                           & $<4.6\times$10$^{-12}$    & --   & 3$\sigma$\,UL\\
\sw/XRT PC$^{e}$ &00032370004  &2012-08-28 07:21:39  &2012-08-28 07:29:56 &      484    & $^{d}$                          & $^{d}$                          & $<2.3\times$10$^{-12}$    & --   & 3$\sigma$\,UL\\
 \noalign{\smallskip} 
 \sw/XRT PC$^{f}$ &  All 3$\sigma$\,UL   &  2011-10-15 19:16:28    &   2012-08-28 07:29:56 &  6519 &   $^{d}$   &   $^{d}$       & $<1.1\times$10$^{-12}$ & --   & 3$\sigma$\,UL   \\

 \noalign{\smallskip} 

{\it XMM} Epic-pn$^{e}$    &0764191301  &2016-03-06 09:21:41 &2016-03-06 19:34:01 &30678            & $0.77_{-0.47}^{+0.70}$    &  $-0.11_{-0.26}^{+0.28}$        & $(5.2^{+0.8}_{-0.3})\times$10$^{-13}$& -- & 29.6/19$^{g}$  \\ 
{\it XMM} Epic-MOS$^{e}$ &0764191301  &2016-03-06 08:58:52 &2016-03-06 19:38:06 &35923            &        --                            &            --                                  &           --                          & -- &  -- \\
\noalign{\smallskip}
  \hline
  \end{tabular}
  \end{center}
  \tablefoot{A power-law was used to describe  the BAT data, while an absorption component has been included in the case of the XRT and \xmm\ data. 
$^a$:  Observed.    $^b$:  $N_{\rm H}<5.1\times10^{22}$\,cm$^{-2}$ 3\,$\sigma$ c.l..    $^c$:  Cash statistics.   $^d$:  Adopting the same spectral parameters as in observation  00505516000. 
  $^e$:  Serendipitous observations.   $^f$:  Combination of all data after T$+10^4$\,s (from 00036122001 to 00032370004). 
   $^g$:  Simultaneous EPIC-pn and MOS1 fit. }
  \end{table*}

In this paper, we present the newly collected XRT and UVOT 
monitoring data of the three sources  
2XMM~J185114.3$-$000004, IGR~J17407$-$2808, and IGR~J18175$-$2419, 
which showed an X--ray activity reminiscent of what is typically observed from the SFXTs during either 
the BAT triggers or the follow-up observations in the soft X--rays. 
Our main goal is to use the new data to investigate the associations of these systems with the SFXT class. 
We also supplement our data set by including: 
{\it (i)} serendipitous archival \xmm\ observations of \igr;  
{\it (ii)} serendipitous archival \xmm\ and ESO Very Large Telescope (VLT) observations of \xmmj;
{\it (iii)} archival \inte\ observations of \nosrc\ 
carried out with the IBIS/ISGRI instrument.

        \section{The sample \label{sfxt3cand:sample}} 

\xmmj\ is a source in the \xmm\ XMMSSC catalogue 
\citep[][]{Watson2009:2XMMserendip,Lin2012:XMMSerendip} that 
triggered the BAT on 2012 June 17 \citep[][]{Barthelmy2012:GCN13367}. 
At the time of discovery, the source showed an increase in the X--ray flux that was at least a factor of 
40 compared to previous detections. 
Recently, \citet[][]{Bamba2016:2XMM185114} have analysed a $\sim 100$\,ks \suzaku\
observation of the supernova remnant (SNR) G32.8$-$0.1 which serendipitously included \xmmj\ 
and found evidence of 
high time variability with flares on timescales of a few hundred seconds superimposed on a
general decaying X--ray flux during the observation. No pulsations were found, but the flares
were noticed to be spaced by $\sim 7000$\,s from one another. The 3--10\,keV spectrum
was characterised by a high absorption ($N_{\rm H}\sim10^{23}$\,cm$^{-2}$) and a photon index 
$\Gamma\sim 1.6$, with a 2--10\,keV flux of $\sim10^{-11}$~erg\,cm$^{-2}$\,s$^{-1}$.

\igr\ was discovered with \inte\ as a new transient on 2004 Oct 9  
\citep[][]{Gotz2004:gcn2793,Kretschmar2004:atel345} and 
associated with either SBM2001 10 \citep[][]{Sidoli2001:catgcr} 
or the ROSAT source 2RXP~J174040.9$-$280852.   
Based on its hard X--ray behaviour, 
\citet[][]{Sguera2006} proposed it is a candidate SFXT 
as they noted peculiarly quick strong flares 
(20--60\,keV, peak flux 800\,mCrab) lasting a couple of minutes. 
\citet[][]{Heinke2009:chandra} found a likely association with the soft X--ray source
 CXOU~J174042.0$-$280724 \citep[][]{Tomsick2008:chandra}. 
Following a BAT trigger on 2011 October 15, \citet[][]{Romano2011:atel3685} identified 
CXOU~J174042.0$-$280724 as the soft X--ray counterpart of \igr. This in turn led 
to the identification of the candidate IR counterpart \citep[][]{Greiss2011:atel3688,Kaur2011:atel3695}. 
If confirmed, this could disproof the SFXT hypothesis, as the most likely optical companion 
seems at present a late type F dwarf. 

The transient \nosrc\ was serendipitously discovered \citep[][]{Grebenev2013:18175_2419} 
in the IBIS/ISGRI data of the \inte\ observations performed in the direction of the X--ray nova 
SWIFT~J174510.8$-$2624 on 2012 September 26. 
The characteristic short (1\,h) flare displayed by the source at discovery, combined with 
its spectrum described by a hard power-law ($\Gamma\sim2.1$) and a possible exponential cut-off above 
80\,keV, suggested \nosrc\ could be a newly discovered SFXT source. 
No further detections of the source have been reported to date. 

\setcounter{table}{2}  
 \begin{table} 	
\tabcolsep 4pt   
\begin{center} 	
 \caption{Log of X--ray observations of  IGR~J18175$-$241. } 	
 \label{sfxt3cand:tab:18175log} 	
 \scriptsize
\begin{tabular}{ rrrrr } 
 \hline 
 \hline 
 \noalign{\smallskip} 
 Instrument  &  ObsID          & Start time    & End time    & Expo.    \\ 
                    &           & (UT)  & (UT)  &(s)         \\
 \noalign{\smallskip} 
 \hline 
 \noalign{\smallskip} 
\sw/XRT &00034136001		&	2016-03-06 12:41:58	&	2016-03-06 12:57:54	&	955	\\
\sw/XRT &00034136002		&	2016-03-07 07:58:45	&	2016-03-07 08:16:53	&	1088	\\
\sw/XRT &00034136003		&	2016-03-08 09:31:28	&	2016-03-08 09:47:53	&	963	\\
\sw/XRT &00034136004		&	2016-03-09 00:03:16	&	2016-03-09 00:16:53	&	43	\\
\sw/XRT &00034136005		&	2016-03-10 22:12:58	&	2016-03-10 22:27:56	&	888	\\
\sw/XRT &00034136006		&	2016-03-11 14:05:57	&	2016-03-11 14:21:54	&	958	\\
\sw/XRT &00034136007		&	2016-03-12 07:43:28	&	2016-03-12 07:59:54	&	985	\\
\sw/XRT &00034136008		&	2016-03-13 01:04:33	&	2016-03-13 01:20:54	&	963	\\
\sw/XRT &00034136009            &     2016-03-14 04:11:53     &       2016-03-14 04:27:53    &       20      \\
\noalign{\smallskip}
  \hline
  \end{tabular}
  \end{center}
  \end{table} 

 \setcounter{table}{3}  
\begin{table}
 \tabcolsep 4pt   
\begin{center}
  \caption{Optical data on 2XMM~J185114.3$-$000004: VLT/NACO and \sw/UVOT observations.  
  \label{sfxt3cand:tab:2xmmlog_opt} }
\scriptsize
\begin{tabular}{ccccc}
\hline 
 \hline 
 \noalign{\smallskip}  
{\bf Instrument}  &	Time mid observation &  Exposure		    & Filter & Mag               \\
                   &	      (UT)        &  (s)			    &		 &	                \\ %
 \noalign{\smallskip}  
\hline
 \noalign{\smallskip}  
VLT/NACO          & 2012-07-12 07:41:28 &  $5 \times 2 \times 60$ s  & $J$    & $15.8 \pm 0.1$    \\
VLT/NACO          & 2012-07-12 07:54:45 &  $5 \times 2 \times 60$ s  & $Ks$   & $11.7 \pm 0.1$    \\
\sw/UVOT & 2012-06-26 06:57:11 &  8890  & $u$       &  $>21.84$   \\
\sw/UVOT & 2012-06-17 16:49:25 &    533  & $b$       &  $>20.56$   \\
\sw/UVOT & 2012-06-17 20:44:42 &  1834  & $v$        &  $>20.21$  \\
\sw/UVOT & 2012-09-04 16:58:10 & 15714 & $m2$    &  $>22.11$ \\
\sw/UVOT & 2012-07-13 17:33:49 &   8589 & $w1$     &  $>21.79$ \\
\sw/UVOT & 2012-06-24 09:58:43 &   4012 & $w2$     &  $>21.62$ \\
 \noalign{\smallskip}  
\hline
\end{tabular}
\end{center}
\tablefoot{Magnitudes are in the Vega system and not corrected for Galactic extinction.}
 \end{table}

        \section{Data reduction \label{sfxt3cand:data}} 

The \sw\ and \xmm\
observing logs for 2XMM J185114.3$-$000004, IGR J17407$-$2808,  and IGR J18175$-$2419 are reported in 
Table~\ref{sfxt3cand:tab:2xmmlog}, \ref{sfxt3cand:tab:17407log}, and 
\ref{sfxt3cand:tab:18175log}, respectively. The VLT observation logs are in 
Table~\ref{sfxt3cand:tab:2xmmlog_opt}.

        \subsection{{\em Swift} \label{sfxt3cand:data_swift}} 

The \sw\ data were uniformly processed and analysed using the standard software 
({\sc FTOOLS}\footnote{\href{https://heasarc.gsfc.nasa.gov/ftools/ftools_menu.html}{https://heasarc.gsfc.nasa.gov/ftools/ftools\_menu.html.  } } 
v6.18), calibration (CALDB\footnote{\href{https://heasarc.gsfc.nasa.gov/docs/heasarc/caldb/caldb_intro.html}{https://heasarc.gsfc.nasa.gov/docs/heasarc/caldb/caldb\_intro.html.}}  20160113), and methods. 
In particular, background-subtracted \sw/BAT light curves were created in the standard energy bands 
and mask-weighted spectra were extracted during the first orbit of the first automated target (AT) observation. 
We applied an energy-dependent systematic error vector to the BAT data. 
The \sw/XRT data were processed and filtered with the task {\sc xrtpipeline} (v0.13.2). 
Pileup was corrected, when required, by adopting standard procedures 
\citep[][]{vaughan2006:050315,Romano2006:060124}. In these cases, the size of the point spread 
function (PSF) core affected by pile-up was determined by comparing the observed versus the nominal PSF, 
and excluding from the analysis all the events that fell within that region. 
In the case of \xmmj\  (Table~\ref{sfxt3cand:tab:2xmmlog}),  
we extracted source events from annuli with inner/outer radii of  4/20 pixels  
(note that for XRT one pixel corresponds to $2.36$\arcsec) during the first observation. In all other 
cases a circle with a radius of 20 pixels was adopted. The background events were extracted  
from an annular region with an inner radius of 80 pixels and an external radius of 120 pixels 
centered at the source position. 
For \igr\ (Table~\ref{sfxt3cand:tab:17407log}),  
the source events were extracted from annuli with inner/outer radii of 5/20 pixels in the first observation, 
and a circular region with a radius of 20 pixels in all other cases. Background events were extracted from annuli
with inner/outer radii of 60/120 pixels centered at the source position. 
The XRT light curves were corrected for PSF losses vignetting 
by using the {\sc xrtlccorr} tool and were background subtracted.  
In all observations where no detection was achieved, the corresponding 3\,$\sigma$ upper limit 
on the X--ray count rate was estimated by using  the tasks {\sc sosta} and {\sc uplimit} within {\sc XIMAGE}  
(with the background calculated in the neighbourhood of the source position) 
and the Bayesian method for low counts experiments adapted from \citet[][]{KraftBurrowsNousek1991}. 
For our spectral analysis, we extracted events in the same regions as those adopted to create the 
light curve; ancillary response files were generated with  the task {\sc xrtmkarf} 
to account for different extraction regions, vignetting, and PSF corrections. 

The \sw/UVOT observed the targets simultaneously with the XRT.
It used all filters during AT observations and with the `Filter of the Day', 
i.e.\ the filter chosen for all observations to be carried out during a specific day in 
order to minimize the filter-wheel usage, during all other observations. 
The data analysis was performed using the {\sc uvotimsum} and 
{\sc uvotsource} tasks included in {\sc FTOOLS}. The {\sc uvotsource} 
task calculates the magnitude of the source through aperture photometry within
a circular region centered on the best source position and applies the required corrections 
related to the specific detector characteristics. 
We adopted a circular region with a radius of 5\,\arcsec for the photometry of the 
different sources. The background was evaluated in all cases by using circular regions 
with a radius of 10\,\arcsec.

        \subsection{{\em \xmm} \label{sfxt3cand:data_xmm}} 

The \xmm\ EPIC-pn \citep{Struder2001:epic_pn} 
and EPIC-MOS \citep{Turner2001:epic_mos} observations of 
\xmmj\  and \igr\  were processed by using the \xmm\ Science Analysis Software (SAS, v.\ 
15.0)\footnote{\href{http://xmm.esac.esa.int/sas/}{http://xmm.esac.esa.int/sas/.}   }.  

\xmmj\ was serendipitously observed by \xmm\ 3 times (see Table~\ref{sfxt3cand:tab:2xmmlog})
on 2003 October 5 (ObsID 0017740401), on 2003 October 20 (ObsID 0017740501), 
and on 2012 March 18 (ObsID 0671510101). 
In all cases, the source was located at the very rim of the three EPIC cameras field of view (FOV). 
\xmm\ observation data files (ODFs) for \xmmj\ were processed to produce calibrated
event lists using the standard \xmm\ SAS. 
We used the {\sc epproc} and {\sc emproc} tasks to 
produce cleaned event files from the EPIC-pn and MOS cameras, respectively. 
EPIC-pn and EPIC-MOS event files were extracted in the 0.5--10\,keV 
energy range and filtered to exclude high background time intervals.  
The obs.\ 0671510101 was moderately affected by flaring background episodes. 
The cleaned effective exposure time was of 40.8\,ks. 
The obs.\ 0017740401 was not affected by a high background, and thus we retained 
for the following analysis the entire exposure time available (21.4\,ks for the EPIC-pn 
and 27.4\,ks for the two MOS). 
In obs.\ 0017740501, cleaning for the high level background resulted in an 
effective exposure time of 17.9\,ks for the EPIC-pn and 21.2\,ks for the two MOS cameras. 
Source and background spectra were extracted by using regions in the same CCD.  

\igr\ was also observed serendipitously with \xmm\ (see Table~\ref{sfxt3cand:tab:17407log}) 
on 2016 March 6 (ObsID: 0764191301, PI.\ G.\ Ponti) during a Galactic center lobe scan performed 
as an extension of the \xmm\ scan \citep[][]{Ponti2015a,Ponti2015b}.
The source was located at an off-axis angle of about 7.5\,arcmin from the aim point of all 
EPIC cameras, which were operating in full-frame mode using the medium filter. 
We removed an interval of increased background flaring activity at the end of the observation, yielding an effective exposure of 
34.9, 36.4, and 36.3\,ks for the pn, MOS1, and MOS2, respectively. 
Unfortunately, the source was located right at the edge of CCD1 in both MOS cameras and only a small fraction of the point spread 
function was covered, leading to large uncertainties in the flux reconstruction. 
For the MOS2 the uncertainty was too large to be useful 
for any scientific analysis, and thus we discarded these data. The EPIC-pn and MOS1 energy spectra and time series of the source and 
the background were extracted from circular regions. The radii of these regions were chosen to maximize the S/N by 
using the SAS tasks {\tt eregionanalyse} and  {\tt especget}.
We used single and double-pixel events for the EPIC-pn camera and single to quadruple-pixel events for the MOS1. 
For the energy spectra, events with {\tt FLAG $\neq$ 0} were discarded before binning the data to have $S/N\geq 5$ 
in each bin. 
To produce a background-subtracted X--ray light curve of \igr\ 
we selected single- and double-pixel events from the EPIC-pn 
camera in the 0.2--10.0 keV energy band and used a time binning of 200\,s.

        \subsection{{\em ESO VLT} \label{sfxt3cand:data_vlt}} 

We observed the field of \xmmj\ with the  
ESO VLT equipped with NAos COnica (NACO), the adaptive optics (AO) NIR imager and 
spectrometer mounted at the VLT UT4 telescope, in the $J$ and $Ks$ bands. 
Observations were carried out on 2012 July 12 starting at 07:36:01.548 UT 
(see Table~\ref{sfxt3cand:tab:2xmmlog_opt}). 
We used the S27 camera, which has a pixel size of $0.027\arcsec$ and a FOV of 
$28\arcsec \times 28\arcsec$. The visual dichroic element and wavefront sensor were used. 
Image reduction was carried out using the NACO pipeline data reduction, part of the 
ECLIPSE\footnote{\href{http://www.eso.org/projects/aot/eclipse/}{http://www.eso.org/projects/aot/eclipse/.}} package. 
Unfortunately, the observations were affected by natural seeing in excess of $1\arcsec$, with a resulting poor resolution.
Astrometry was carried out by using the  2MASS\footnote{\href{http://www.ipac.caltech.edu/2mass}{http://www.ipac.caltech.edu/2mass/.}}  
catalogues as reference. 
Aperture photometry was performed with the {\sc PHOTOM} task of the {\sc Starlink}  
package\footnote{\href{http://starlink.eao.hawaii.edu/starlink}{http://starlink.eao.hawaii.edu/starlink.}}. 
The photometric calibration was done against the 2MASS catalogue.

\begin{figure}[t]
\vspace{-1.25truecm}
\hspace{-0.5truecm}
  \centerline{\includegraphics[width=10cm,height=8.5cm,angle=0]{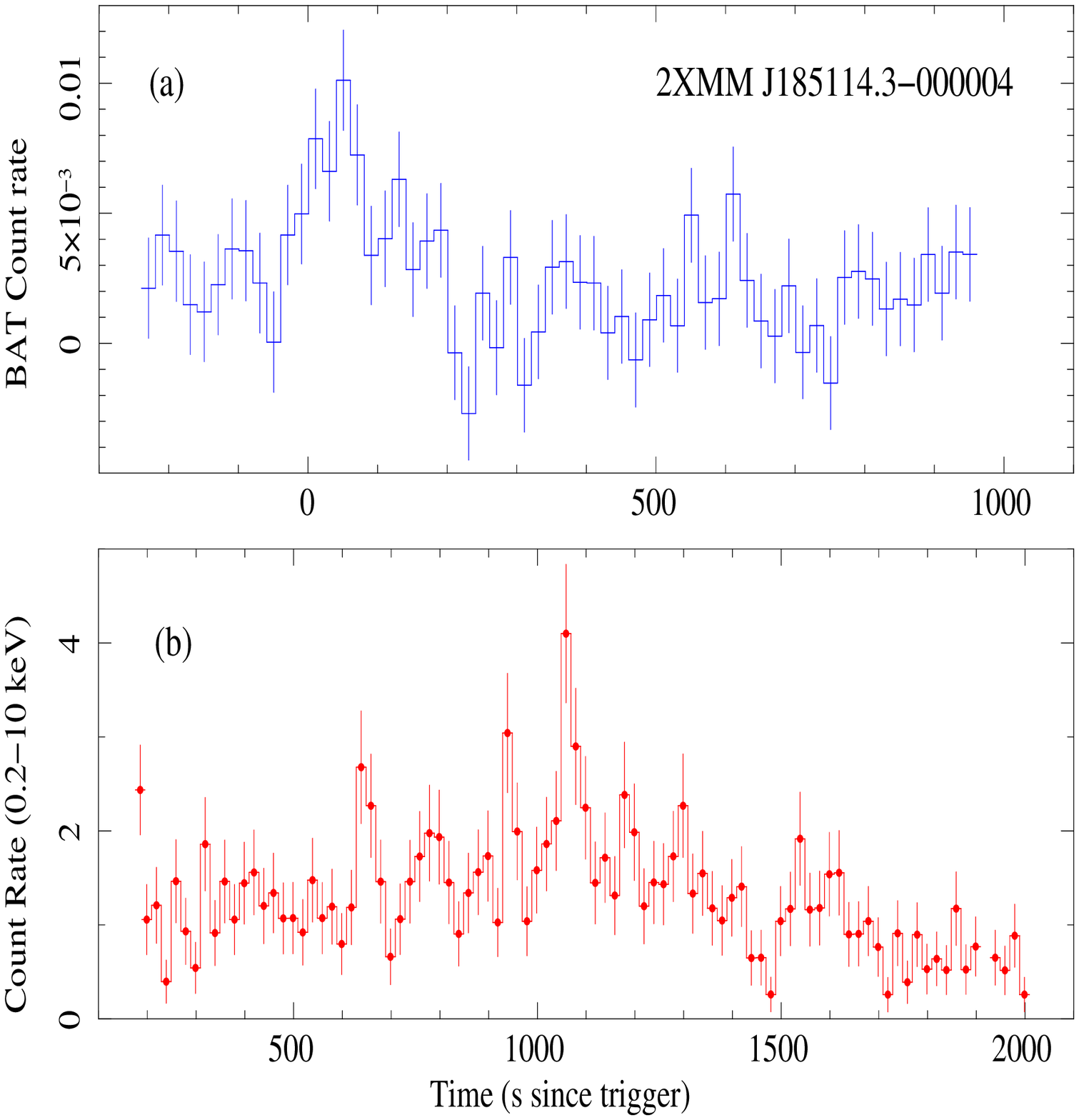}}
\vspace{-1.8truecm}
\caption{Light curves of the outburst in 2012 June 17 of \xmmj\ (first \sw\ orbit data). 
{\bf a)} BAT light curve in the 14--50 keV with a time binning of 20 s. 
{\bf b)} XRT light curve in the 0.2--10keV, rebinned to have at least 10 counts bin$^{-1}$. 
Note the different x-axis scales. }    
\label{sfxt3cand:fig:2xmm_batxrtlcv} 
\end{figure}

        \subsection{Other data \label{sfxt3cand:data_rest}} 

\inte\ data were only used for the source IGR J18175$-$2419 and the corresponding results are described in 
Sect.~\ref{sfxt3cand:data_18175}.

        \section{Results \label{sfxt3cand:results}}

        \subsection{\xmmj \label{sfxt3cand:data_2xmm}}

%
%
The source  2XMM J185114.3$-$000004  triggered the BAT on 2012 June 17 at  $T_0$ $=$ 15:46:55 UT 
\citep[64\,s image trigger=524542, ][]{Barthelmy2012:GCN13367}, resulting in a 7\,$\sigma$ detection. 
\sw\ performed an immediate slew to the target and XRT started observing at $T_0+$172\,s. 
The AT (sequence 00524542000-001) ran for seven orbits, until $T_0+17.5$\,ks.  
Follow-up observations were obtained daily (sequences 00524542002--010). 
Additional ToO observations were first performed when the source rebrightened few days later 
(PI P. Romano, sequences 00524542011--017), 
and then also on August 8--31 of the same year (PI P. Romano, sequences 00032512001--005). 
The \sw\ data therefore cover the first 18\,d after the beginning of the outburst, and then  
about three weeks more later that year (see Table~\ref{sfxt3cand:tab:2xmmlog}).  
The source is only detected in 5 observations (00524542000--1 and 9, 00032512001 and 5).  
We also found an archival \sw\ serendipitous observation performed on 2012 November 20 
(00044565001) which also resulted in a non-detection of the source. 

We used 4\,ks of the XRT/PC mode data collected during the outburst in 2012 June 17 
and the simultaneous \sw/UVOT images to 
obtain an astrometrically-corrected source position 
\citep[see][]{Evans2009:xrtgrb_mn, Goad2007:xrtuvotpostions} at: 
RA(J$2000)=18^{\rm h}\, 51^{\rm m}\, 14\fs50$, 
Dec(J$2000)=-00^{\circ}\, 00^{\prime}\, 04\farcs1$ (90\,\% confidence level, c.l., uncertainty of 1\farcs7).  
This position is $1\farcs7$ from the catalogued position of \xmmj.        
It is also $0\farcs6$ from the Two Micron All Sky Survey  
source 2MASS~J18511447$-$0000036 \citep[][]{Skrutskie2006:2MASS}.   

\begin{figure}[t]
  \centerline{\includegraphics[width=5.5cm,height=9.5cm,angle=-90]{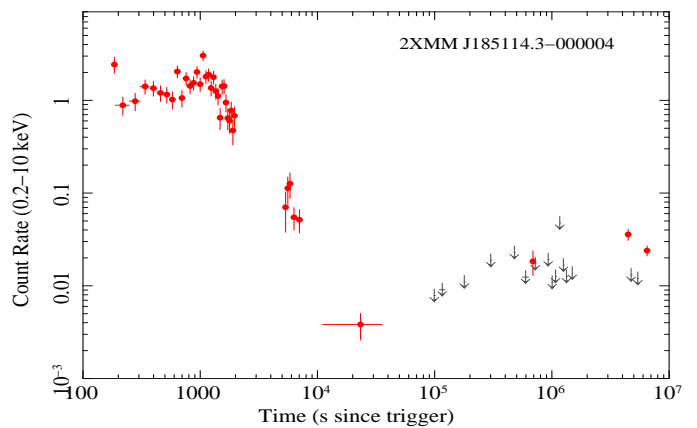}}
\caption{\sw/BAT and XRT light curve obtained from the entire XRT dataset on \xmmj. Grey downward-pointing arrows 
correspond to the 3\,$\sigma$ upper limits.}    
\label{sfxt3cand:fig:2xmm_xrtlcv} 
\end{figure}

Figure~\ref{sfxt3cand:fig:2xmm_batxrtlcv} shows the \sw\ BAT (14--50\,keV)
and XRT (0.2--10\,keV) light curves of the source extracted from the first orbit of data collected 
at the beginning of the 2012 June outburst, 
while Fig.~\ref{sfxt3cand:fig:2xmm_xrtlcv} shows the light curve derived from the 
whole XRT dataset. 
The XRT light curve reached a maximum of 4.1 counts s$^{-1}$. 
This corresponds 
to an approximate flux of $7.9\times10^{-10}$ erg\,cm$^{-2}$\,s$^{-1}$ (we used  
the conversion factor derived from the fit to the XRT data in observation 00524542000 PC mode,
see Table~\ref{sfxt3cand:tab:2xmmlog}). The lowest X--ray flux from the source was recorded  
at $3.8\times10^{-3}$ counts s$^{-1}$ during the observation 00524542001, thus resulting in an XRT dynamical 
range $\gtrsim 10^{3}$.

The BAT spectrum of the source extracted from observation 00524542000 was fit in the 14--70\,keV 
energy range with a simple power law (see Table~\ref{sfxt3cand:tab:2xmmlog}). 
The XRT spectrum extracted from observation 00524542000 was fit in the 0.3--10\,keV energy range
with an absorbed power law  (see Table~\ref{sfxt3cand:tab:2xmmlog}). 
We then considered the total BAT spectrum and the nearly simultaneous XRT spectrum for a broad band fit. 
Factors were included in the fitting to allow for the different exposures of the two spectra, 
normalisation uncertainties between the two instruments (generally constrained within $\sim 10\,\%$), 
and a likely spectral variation throughout the exposure.

\begin{figure}[t]
\hspace{-0.3truecm}
\centerline{\includegraphics[width=5.5cm,height=9.4cm,angle=-90]{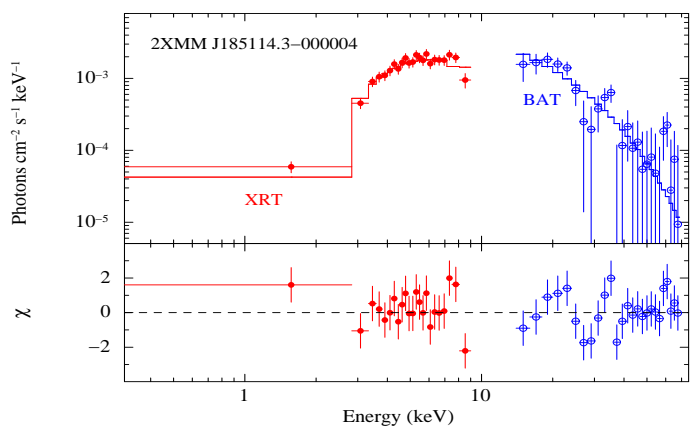}}
\caption{Spectroscopy of the 2012 June 17 outburst of \xmmj. 
The top panel shows simultaneous XRT/PC (filled red circles) and BAT data (empty blue circles) fit with a 
{\sc phabs*cutoffpl} model. The residuals from the best fit are shown in the 
bottom panel (in units of standard deviations). 
}    
\label{sfxt3cand:fig:2xmm_batxrtspec} 
\end{figure}

Several models typically used to describe the X-ray emission from accreting pulsars in HMXBs 
were adopted \citep[e.g.][and references therein]{White1983,Coburn2002,Walter2015:HMXBs_IGR}, 
including:  
{\it i)} an absorbed power law (hereon POW), 
{\it ii)} an absorbed power law with high energy exponential roll-off at an e-folding energy $E_{\rm f}$ 
             ({\sc phabs*cutoffpl} in {\sc XSPEC}, hereon CPL),
{\it iii)} and an absorbed power law with a high energy cut-off at an energy $E_{\rm c}$, and an e-folding energy $E_{\rm f}$ 
             ({\sc phabs*power*highecut}, hereon HCT). 
The results for \xmmj\ are reported in Table~\ref{sfxt3cand:tab:2xmmbroadband_specfits}. 
The PL model clearly yielded unacceptable results. 
The CPL model produced significantly better results and gave 
$N_{\rm H}=(13\pm4)\times$10$^{22}$~cm$^{-2}$, $\Gamma=0.11_{-0.65}^{+0.68}$, 
and $E_{\rm f} =10_{-3}^{+7}$~keV (see Fig.~\ref{sfxt3cand:fig:2xmm_batxrtspec}). 
These values are compatible with those usually expected for highly magnetised accreting NSs,
SFXTs in particular \citep[e.g.][]{Romano2011:fermi11}. 
The HCT model is, on the other hand, unable to constrain the cut-off energy (see Table~\ref{sfxt3cand:tab:2xmmbroadband_specfits}); 
as the addition of one free parameter does not improve the statistics significantly 
($F$-test probability of 0.269 with respect to the CPL model), we favour the CPL model. 

%
The UVOT data obtained simultaneously with the XRT ones only yield 3$\sigma$ upper limits in all filters
(see Table~\ref{sfxt3cand:tab:2xmmlog_opt}). 
This is not surprising, given that the reddening along the line of sight is $E(B-V) \sim19$,  
implying an extinction of $A_V \sim 60$ mag.

 \setcounter{table}{4}  
 \begin{table}
 \tabcolsep 4pt   
\begin{center}
 \caption{Spectral fits of the simultaneous XRT and BAT data of  2XMM~J185114.3$-$000004 
during the outburst on 2012 June 17.  
}
\scriptsize
 \label{sfxt3cand:tab:2xmmbroadband_specfits}
 \begin{tabular}{lllllllll}
 \hline
 \hline
 \noalign{\smallskip}
Model $^{a}$ &$N_{\rm H}$     &$\Gamma$                &E$_{\rm c}$           &E$_{\rm f}$         &$F_{\rm 0.5-100\,keV}^{b}$          &$\chi^{2}$/dof  \\
       & ($10^{22}$ cm$^{-2}$) &                                 & (keV)                   & (keV)                & (erg cm$^{-2}$ s$^{-1}$)   &  \\
   \hline
 \noalign{\smallskip}
\sc{POW}  &$24_{-3}^{+4}$  &$2.1\pm0.4$         &    --                        &    --                          &   $3.8\pm1.1$   &$93.7/67$          \\ 
\sc{CPL}   &$13\pm4$   &$0.11_{-0.65}^{+0.68}$   &    --                      &$10_{-3}^{+7}$            &  $1.7\pm0.2$   &$69.3/66$         \\  
\sc{HCT}  &$11\pm4$   &$0.17_{-0.42}^{+1.24}$   &$5_{-5}^{+16}$         &$10_{-4}^{+5}$            &  $1.7\pm0.3$   &$68.0/65$          \\  
\noalign{\smallskip}
  \hline
  \end{tabular}
  \end{center}
  \tablefoot{$^a$:  POW=absorbed powerlaw. 
                             CPL=power law  with high energy exponential with e-folding energy at E$_{\rm f}$ (keV). 
                             HCT=absorbed power law with a high energy cut-off at E$_{\rm c}$  and e-folding energy E$_{\rm f}$. 
                   $^b$:  Unabsorbed flux in units of $10^{-9}$ erg cm$^{-2}$ s$^{-1}$.    }
  \end{table}

\begin{figure}[t] 
\vspace{-1.2truecm}
\centerline{\includegraphics[width=9cm,angle=0]{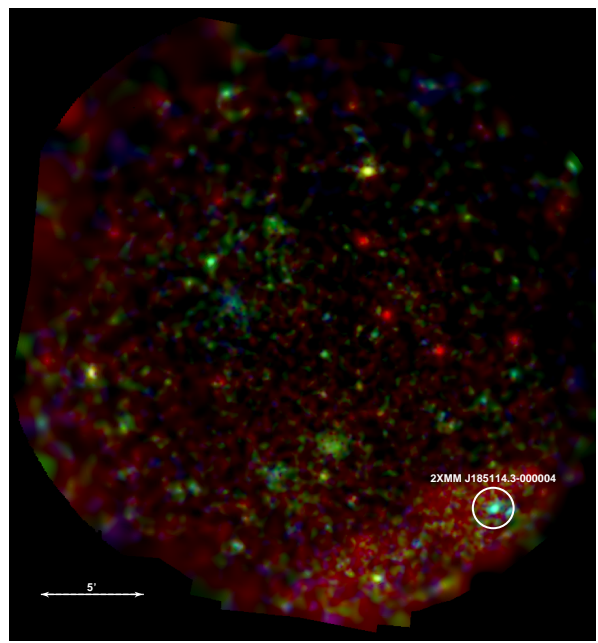}}
\vspace{-1truecm}
\caption{\xmm\ FOV of the observations 0017740401 and 0017740501, 
detector background subtracted, vignetting corrected, combining 
EPIC-pn and MOS. Red, green, blue correspond to 0.5--2.0\,keV, 2.0--4.5\,keV and
4.5--12\,keV. The diffuse emission around \xmmj\ is due to SNR G32.8$-$0.1. 
}    
\label{sfxt3cand:fig:2xmmimg} 
\end{figure}

\begin{figure}[t]
\centering
\includegraphics[scale=0.33, angle=-90]{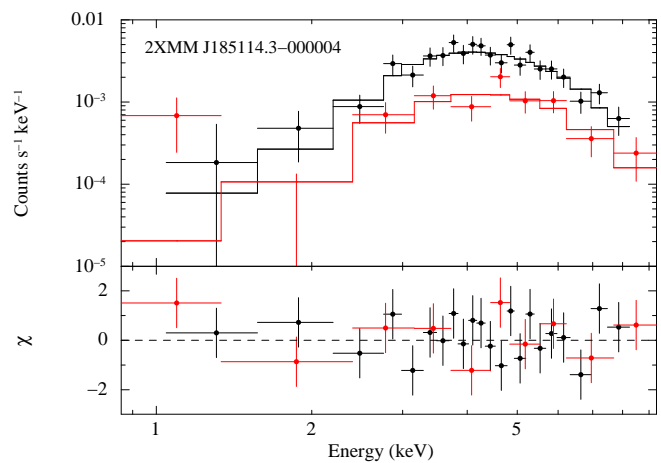}
\caption{Merged \xmm\ EPIC-pn+MOS1+MOS2 spectra extracted from the observations 0017740401 (black) and 0017740501 (red). 
The best fit model is obtained with an absorbed power law. The residuals from the fit are shown in the bottom panel.}    
\label{sfxt3cand:fig:2xmmspectra} 
\end{figure}

By using our ESO VLT observations, 
inside the XRT error circle we detect a single source at the following position: 
RA(J$2000)=18^{\rm h}\, 51^{\rm m}\, 14\fs48$, 
Dec(J$2000)=-00^{\circ}\, 00^{\prime}\, 03\farcs6$ (90\% c.l.\ associated uncertainty $0\farcs3$). 
At the epoch of our observation, we measure for this source 
$J = 15.8 \pm 0.1$ mag and $Ks = 11.7 \pm 0.1$ mag (Vega system; not corrected for Galactic extinction, 
see Table~\ref{sfxt3cand:tab:2xmmlog_opt}). 
This source is present in the 2MASS catalogue (2MASS~J18511447$-$0000036), with magnitudes 
$J > 15.6$ mag, $H = 13.23 \pm 0.07$ mag, and $K = 11.80 \pm 0.04$ mag, in agreement with our measurements.

\xmmj\ was detected by the EPIC-pn at an average count-rate (0.5--10\,keV) of (9.4$\pm$0.7)$\times$10$^{-3}$~count s$^{-1}$ 
in obs.\ 0017740401 and (3.5$\pm$0.5)$\times$10$^{-3}$~count s$^{-1}$ in obs.\ 0017740501. 
The corresponding EPIC-pn spectrum could be well fit in both cases with an absorbed power-law model 
(see Table~\ref{sfxt3cand:tab:2xmmlog}). 

Given the relatively low count-rate of the source, we combined in each of the two observations the EPIC-pn and MOS spectra 
by following the on-line SAS data analysis 
threads\footnote{See 
http://xmm.esac.esa.int/sas/current/documentation/threads/ \\ 
epic\_merging.shtml.}.
The resulting spectra (shown in Fig.~\ref{sfxt3cand:fig:2xmmspectra}) provided values for the best fit parameters  
in agreement with those estimated above by using only the 
EPIC-pn data (to within the uncertainties).    

In obs.\ 0017740401, where the statistics was better, we also inspected the source light curve and event file 
searching for timing features. However, the statistics was far too poor to allow for a meaningful timing analysis. 

The source was not detected in the obs.\ 0671510101. From the EPIC-pn data we estimated a 
3\,$\sigma$ upper limit on the source count-rate of 8$\times$10$^{-4}$ in the 0.5--10\,keV 
energy range. Assuming the same spectral shape as in obs.\ 0017740501, the count-rate upper limit would 
translate into a flux of $F_{0.5-10~{\rm keV}}<$4$\times$10$^{-15}$~erg~cm$^{2}$~s$^{-1}$.

\begin{figure}[t]
\vspace{-1.4truecm}
\hspace{-0.5truecm}
 \centerline{\includegraphics[width=10cm,height=8.5cm,angle=0]{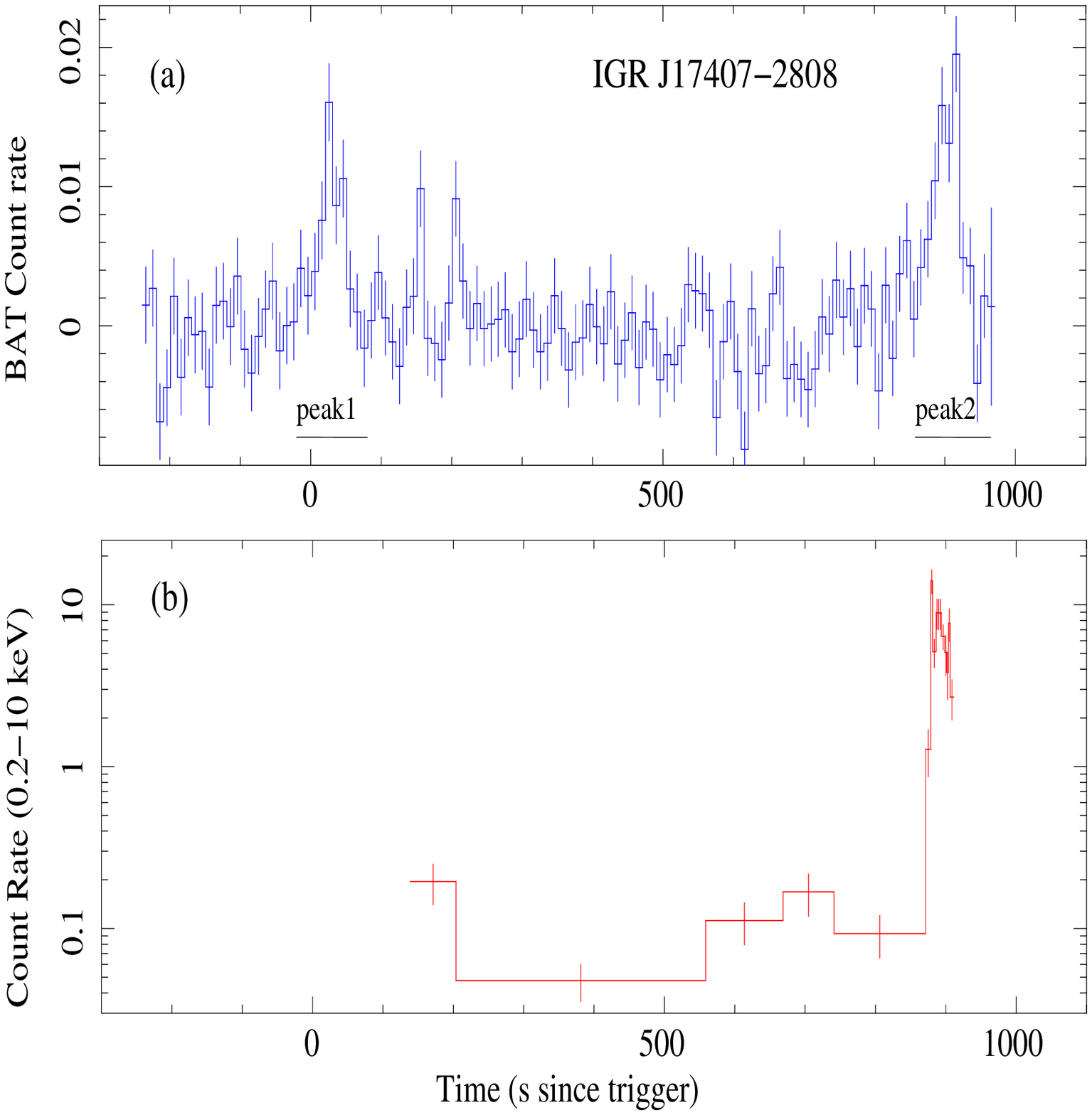}}
\vspace{-1.8truecm}
\caption{Light curves of the 2011 October 15 outburst of \igr\ 
(first \sw\ orbit data). 
{\bf a)}  BAT light curve in the 14--50~keV energy band and a binning of 10~s. 
The horizontal lines mark the time intervals used for the spectral extraction (peak1 and peak2). 
 {\bf a)}  The XRT light curve in the 0.2--10~keV energy band. An adaptive 
binning has been used to achieve in each point a signal-to-noise ratio of S/N=3.}    
\label{sfxt3cand:fig:17407_batxrtlcv} 
\end{figure}

        \subsection{\igr \label{sfxt3cand:data_17407}}

%
%
%
The source \igr\  triggered the BAT on 2011 October 15 at $T_0$ $=$ 01:12:40 UT 
\citep[image trigger=505516, ][]{Romano2011:atel3685}.   
\sw\ immediately slewed to the target and XRT 
started observing at $T_0+$131\,s. 
The AT (sequence  00505516000) ran for two orbits until $T_0+6.6$\,ks. However, 
due to a loss of lock of the star tracker, only the first orbit has a stable attitude and 
can be used for scientific analysis. Follow-up pointings toward the source were obtained daily 
as target of opportunity  (ToO) observations (PI P. Romano, sequences 00036122001--007,
see Table~\ref{sfxt3cand:tab:17407log}). 
The source was only detected during the AT observation (00505516000), but never in the 
following monitoring campaigns and in the serendipitous observations found in the \sw\ archive   
(see Table~\ref{sfxt3cand:tab:17407log}).

We used 772\,s of XRT/PC mode data and the simultaneously collected UVOT images to 
obtain the best astrometrically-corrected source position 
at: 
RA(J$2000)=17^{\rm h}\, 40^{\rm m}\, 42\fs10$, 
Dec(J$2000)=-28^{\circ}\, 07^{\prime}\, 26\farcs0$. The associated  
uncertainty at 90\,\% c.l.\ is 2\farcs4. 
This position is $1\farcs5$ from CXOU~J174042.0$-$280724, so we can confirm the 
association between the two sources as preliminarily reported by \citet[][]{Romano2011:atel3685}. 

\begin{figure}[t]
\centering
\hspace{-0.4truecm}
 \centerline{\includegraphics[width=5.5cm,height=9.3cm,angle=-90]{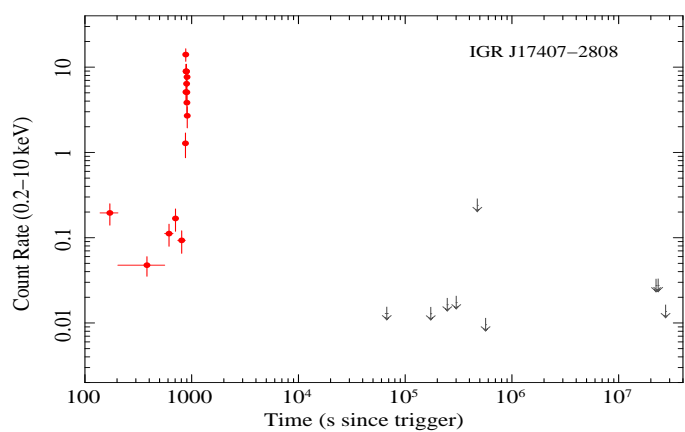}}
\caption{Light curve of the XRT dataset of \igr. 
Grey downward-pointing arrows correspond to the 3\,$\sigma$ upper limits calculated for the source 
non-detections. Points after $10^7$\,s correspond to serendipitous observations.}    
\label{sfxt3cand:fig:17407_xrtlcv} 
\end{figure}

Figure~\ref{sfxt3cand:fig:17407_batxrtlcv} shows the BAT (14--50\,keV)
and XRT (0.2--10\,keV) light curves of the first orbit of data,  
while in Fig.~\ref{sfxt3cand:fig:17407_xrtlcv} we plot the light curve derived from the 
whole XRT dataset. Around $T_0+900$~s, the source displayed a sharp rise in count-rate 
increasing the level of its soft X--ray emission by a factor of 125 in $\sim266$\,s. 
A peak count-rate of about 14 counts s$^{-1}$ is recorded in the light curve binned at S/N=3. 
This corresponds to a flux of $\sim2.3\times10^{-9}$  erg\,cm$^{-2}$\,s$^{-1}$
when using the conversion factor derived from the fit to the XRT data in observation 
00505516000 (see Table~\ref{sfxt3cand:tab:17407log}). 
As the lowest point (obtained from observation 00036122007), 
was a 3\,$\sigma$ upper limit at  $9.5\times10^{-3}$ counts s$^{-1}$, 
the overall dynamical range revealed by XRT is in excess of $\sim1400$.

We extracted two distinct BAT spectra that cover the two peaks in the BAT light curve 
(see Fig.~\ref{sfxt3cand:fig:17407_batxrtlcv}a), i.e. 
from $T_0-20$ to 80\,s (``peak1'') and from $T_0+858$ to 965\,s (``peak2''). 
We fit them with a power law, and the results are reported in Table~\ref{sfxt3cand:tab:17407log}).

The XRT spectrum extracted from obs.\  00505516000 was fit in the 0.3--10\,keV energy range using  
\citet[][]{Cash1979} statistics 
with an absorbed power law. 
We measured an absorption column density $N_{\rm H}=(0.84_{-0.84}^{+2.00})\times$10$^{22}$~cm$^{-2}$ consistent 
with the expected Galactic value in the direction of the source 
\citep[$N_{\rm H}^{\rm Gal}= 0.727\times$10$^{22}$~cm$^{-2}$, ][]{LABS}, and 
a power-law photon index $\Gamma=-0.45_{-0.71}^{+0.86}$, as reported in Table~\ref{sfxt3cand:tab:17407log}.

Since no spectral variations could be detected in either the BAT data or in the XRT data due to the low signal, 
we fit together the BAT peak2 spectrum and the XRT spectrum extracted by using all exposure time available. 
Factors were included in the fitting to allow for normalisation uncertainties between the two instruments, 
the different exposures of the two spectra, and a likely spectral variation throughout the exposure.

\begin{figure}[t]
\centering
\hspace{-0.5truecm}
\centerline{\includegraphics[width=5.5cm,height=9.2cm,angle=-90]{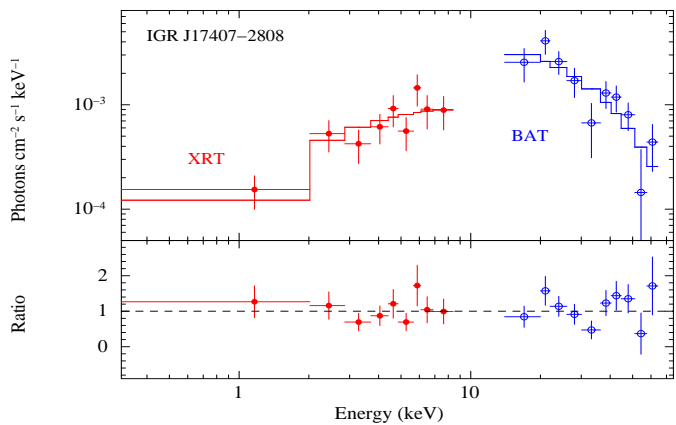}}
\caption{Spectroscopy of the 2011 October 15 outburst of  \igr. 
Top panel: simultaneous XRT/PC data (filled red circles) and BAT data (empty blue circles)  
fit with a {\sc phabs*cutoffpl} model. 
Bottom panel: the ratio between the data and the best fit model. 
}    
\label{sfxt3cand:fig:17407_batxrtspec} 
\end{figure}

The spectra were fit with the models that typically describe either HMXBs 
(see Sect.~\ref{sfxt3cand:data_2xmm}),
or low mass X--ray binaries \citep[LMXB,][]{Done2007,Paizis2006,DelSanto2010:sax1753,Wijnands2015:NSXRB}.  
The results are presented in Table~\ref{sfxt3cand:tab:17407broadband_specfits}.
The POW fit was not acceptable due to large residuals. 
The CPL fit yields 
$\Gamma=-0.76_{-0.53}^{+0.49}$ and E$_{\rm f}=13_{-3}^{+5}$\,keV (see Fig.~\ref{sfxt3cand:fig:17407_batxrtspec}). 
In this fit, we fixed the absorption column density to the value determined from the XRT data alone, i.e.\  
$N_{\rm H}=0.84\times$10$^{22}$~cm$^{-2}$. 
The HCT  model cannot properly constrain the cut-off energy; since the addition of one free parameter does 
not significantly improve the fit ($F$-test probability of 0.423), we favour the CPL model.
A fit with an absorbed blackbody (with $N_{\rm H}$ fixed at 0.84$\times$10$^{22}$~cm$^{-2}$), also 
yielded acceptable results. In this case, the estimated blackbody temperature of about 8\,keV 
would be roughly consistent with that reported previously by \citet[][]{Sguera2006}. 
A fit with an absorbed bremsstrahlung model yields a very high temperature 
and suffers from large systematics in the residuals, so we consider it unacceptable. 
Fits with more sophisticated models, such as 
a comptonisation model ({\sc phabs*(comptt)}) 
and an accretion disk model with multiple blackbody components ({\sc phabs*(diskbb+power)}) 
could not significantly improve the fits. Furthermore, the majority of the spectral fit parameters 
of these models turned out to be largely unconstrained due to the limited statistics of the data.

The source was not detected in any UVOT data obtained simultaneously with XRT. The corresponding 3\,$\sigma$ upper limits 
in all filters were of $u=20.23$, $b=18.61$, $v=17.56$, $m2=20.90$, $w1=20.64$, and $w2=20.33$\,mag 
(Vega system,  not corrected for Galactic extinction). 

\begin{figure}[t]
\hspace{-0.4truecm}
\centerline{\includegraphics[width=5.5cm,height=9.2cm,angle=-90]{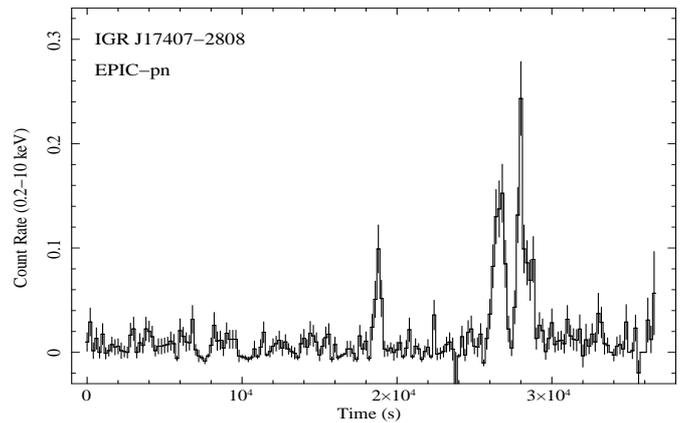}}
\caption{\xmm\ EPIC-pn light curve of \igr\ extracted from the observation 0764191301, with a 200\,s binning.}    
\label{sfxt3cand:fig:17407xmmlcv} 
\end{figure}

\begin{figure}[t]
\hspace{-0.2truecm}
\centerline{\includegraphics[width=5.5cm,height=9.4cm,angle=-90]{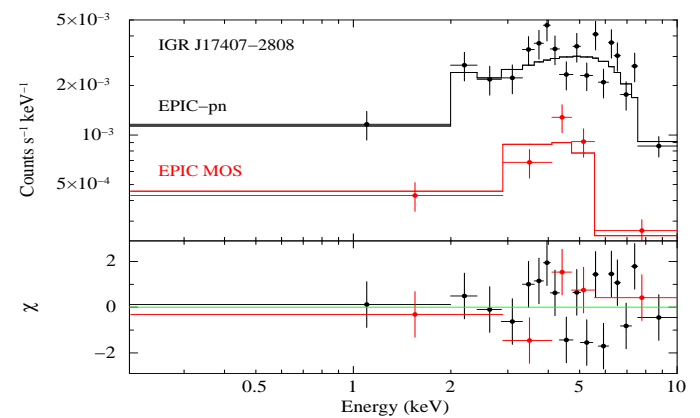}}
\caption{\xmm\ EPIC-pn (black) and MOS1 (red)
spectra extracted from observation 0764191301.  
The best fit is obtained by using an absorbed power-law model. 
The residuals from the fits are shown in the bottom panel.}    
\label{sfxt3cand:fig:17407xmmspectra} 
\end{figure}

\igr\ was also observed serendipitously during an \xmm\ observation performed on 2016 March 6. 
The 0.2--10.0 keV energy band light curve, at a binning of 200\,s, reported in 
Fig.~\ref{sfxt3cand:fig:17407xmmlcv} shows three moderately bright flares, reaching between 0.1 and 0.3 counts\,s$^{-1}$.  
The \xmm\ spectra of the source could be well described  
by using an absorbed power-law model (Fig.~\ref{sfxt3cand:fig:17407xmmspectra}), 
and an $N_{\rm H}=(0.77_{-0.48}^{+0.70})\times$10$^{22}$~cm$^{-2}$, 
consistent with the Galactic value, and with the results obtained with the XRT data. 
The absorbed 0.5--10~keV flux is 5.4$\times$10$^{-13}$~erg cm$^{-2}$ s$^{-1}$,
the lowest X--ray flux measured for this source, enhancing its previously estimated dynamic range (with XRT)
up  to $>4000$.

 \setcounter{table}{5}  
 \begin{table}
 \tabcolsep 4pt   
\begin{center}
 \caption{Spectral fits of the simultaneous XRT and BAT data of IGR~J17407$-$2808
during the outburst on 2011 October 15.  
}
\scriptsize
 \label{sfxt3cand:tab:17407broadband_specfits}
 \begin{tabular}{llllllll}
 \hline
 \hline
 \noalign{\smallskip}
Model $^{a}$ &$N_{\rm H}$          &$\Gamma$                &E$_{\rm c}$                    &E$_{\rm f}$         &$F_{\rm 0.5-100\,keV}^{b}$          &$\chi^{2}$/dof  \\
       & ($10^{22}$ cm$^{-2}$)      &                                  & (keV)                          & (keV)                & (erg cm$^{-2}$ s$^{-1}$)   &  \\
\hline
 \noalign{\smallskip}
\sc{POW}     &$5.3_{-1.7}^{+2.2}$      &$1.6_{-0.4}^{+0.4}$        &    --                          &    --                       &  $6.5\pm1.1$   & $124/93$          \\ 
\sc{CPL}      &$0.84^{c}$                   &$-0.76_{-0.53}^{+0.49}$  &    --                           &  $13_{-3}^{+5}$       &  $3.9\pm0.7$   & $85.3/93$         \\ 
\sc{HCT}     &$0.84_{-0.84}^{+1.87}$  &$-0.45_{-0.68}^{+0.72}$  &$20_{-20}^{+6}$     & $14_{-4}^{+8}$        &  $3.6\pm0.7$   & $83.7/91$         \\ 
\noalign{\smallskip}
 
\hline
 \noalign{\smallskip}
                                      &                       &$kT$                           &                        &           &   \\
                                      &       & (keV)                                  &               & &    &  \\
\hline
 \noalign{\smallskip}
\sc{BB}                            &$0.84$$^{c}$               &$8.7_{-0.5}^{+0.6}$                                        &  --     &    --        &  $3.6\pm0.6$   & $83.4/94$         \\ 
\sc{BRE}                  &$4.5_{-1.4}^{+1.8}$       &$65_{-28}^{+113}$                               &--   &  --                  &  $6.0\pm0.9$   & $114.9/93  $         \\
\noalign{\smallskip}
  \hline
  \end{tabular}
  \end{center}
  \tablefoot{$^a$:  POW=absorbed powerlaw. 
                             CPL=power law  with high energy exponential with e-folding energy at E$_{\rm f}$ (keV). 
                             HCT=absorbed power law with a high energy cut-off at E$_{\rm c}$  and e-folding energy E$_{\rm f}$.  
                             BB=black body with temperature $kT$.   
                             BRE=thermal bremsstrahlung with plasma temperature $kT$. 
     $^b$:  Unabsorbed flux in units of $10^{-9}$ erg cm$^{-2}$ s$^{-1}$.  
     $^c$:  Fixed to the value obtained from the fit to the XRT data (see Table~\ref{sfxt3cand:tab:17407log}). }
  \end{table}

        \subsection{\nosrc \label{sfxt3cand:data_18175}}

%
\nosrc\ was observed by \sw\ as part of our ongoing effort \citep[][]{Romano2015:swift10}
to study SFXTs, candidate SFXTs, 
and classical supergiant HMXBs through long-term monitoring programs with the XRT 
\citep[see][for recent results]{Romano2014:sfxts_paperX}. 
Our monitoring campaign (see Table~\ref{sfxt3cand:tab:18175log}) 
was performed from 2016 March 6 to 14  for 1\,ks a day.  
As the source was poorly known, 
the XRT was set in AUTO mode to best exploit the automatic mode-switching of the instrument 
in response to changes in the observed fluxes \citep{Hill04:xrtmodes}. 
We collected a total of 9 \sw\ observations for a total net XRT exposure time of $\sim 7$\,ks.  
No source was detected within the previously reported INTEGRAL error circle \citep[][]{Grebenev2013:18175_2419}
in either any of the 1\,ks snapshots or in the combined total exposure. We estimated  
a 3\,$\sigma$ upper limit on the source X--ray cout rate of 1.5--3$\times10^{-3}$ counts\,s$^{-1}$, 
which corresponds to 0.8--1.7$\times10^{-13}$ erg cm$^{-2}$ s$^{-1}$ 
(when using, within {\sc PIMMS}, a typical spectral model for SFXTs, 
comprising a power-law with photon index $\Gamma=1.5$ and an 
absorption column density corresponding to the Galactic value in the direction of the source, i.e.\   
$N_{\rm H}^{\rm Gal}=2.66\times10^{21}$ cm$^{-2}$).

\begin{figure*} 
 \centerline{
\includegraphics[width=8.9cm,angle=0]{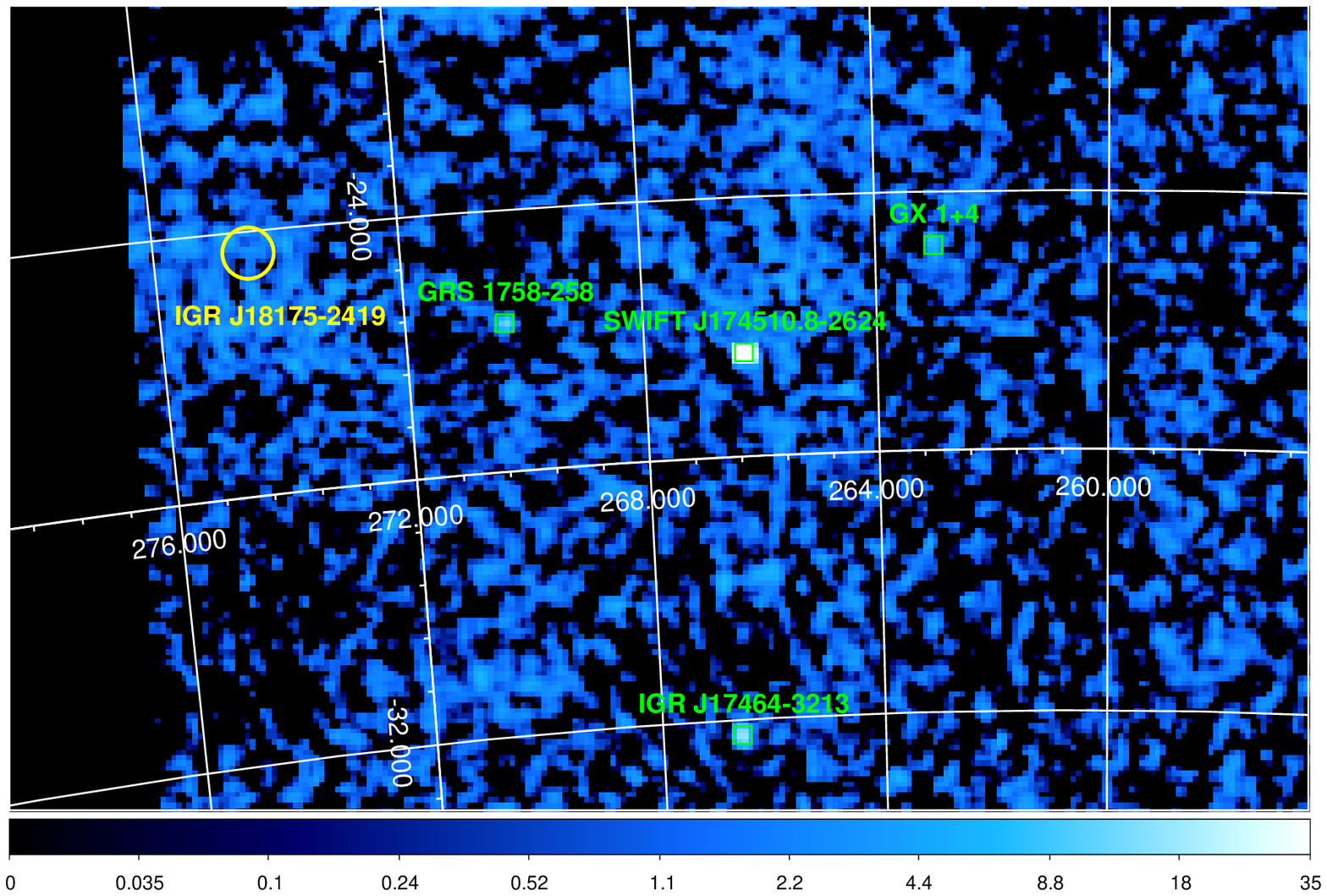}
                  \includegraphics[width=8.7cm,angle=0]{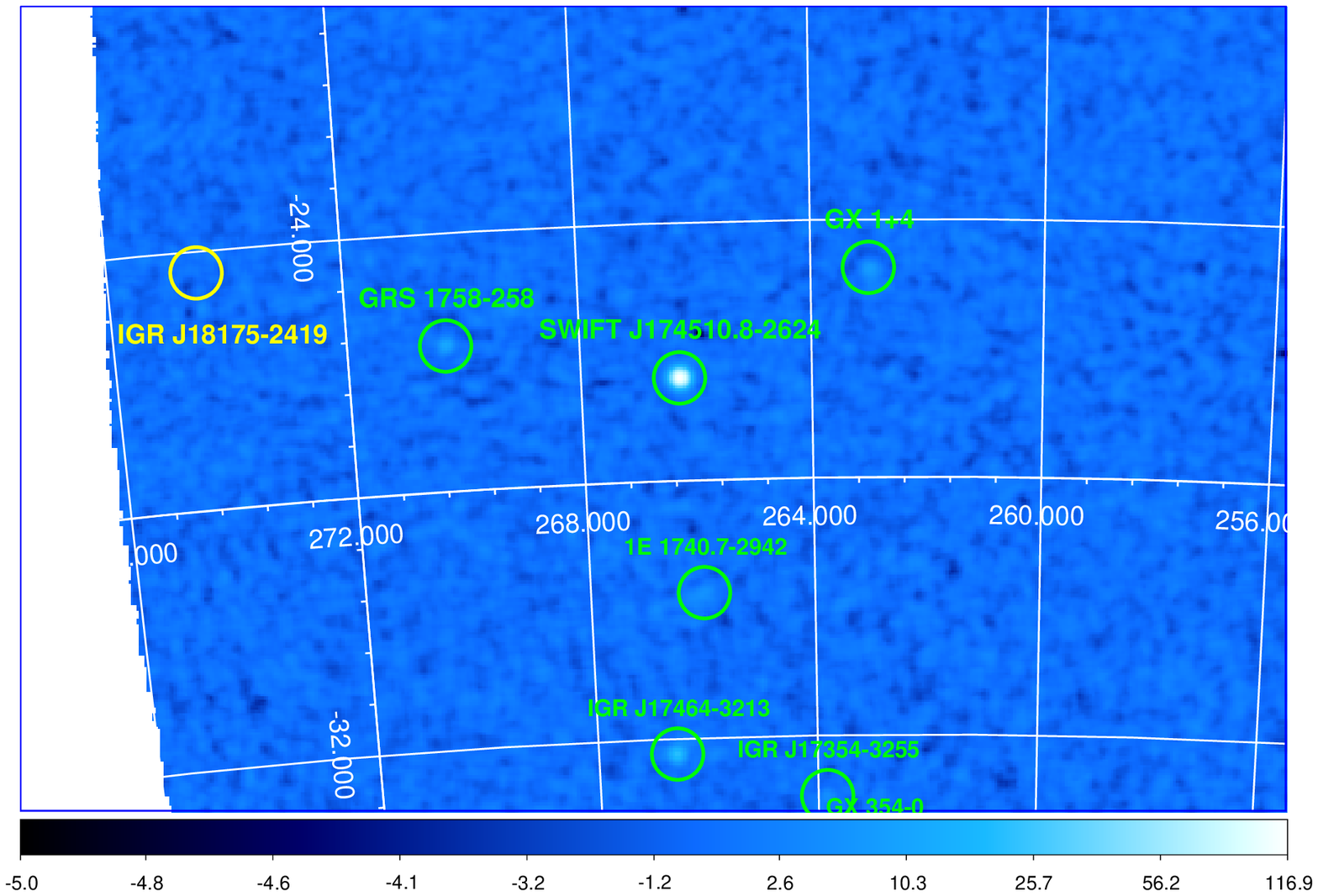}}
  \caption{{\bf Left}: IBIS/ISGRI mosaic extracted with the OSA software from the SCW 51 in revolution 1215 (20--80~\,keV energy band). We marked 
  (green squares) on the mosaic the position of 4 sources detected with a sufficiently high significance 
  (SWIFT~J174510.8$-$2624, GRS~1758$-$258, GX~1$+$4, IGR~J17464$-$3213) and that previously reported for \nosrc. 
  The latter is indicated by using a yellow circle with a radius of  
  24\,$\arcmin$ centered on the source best position provided by \citet{Grebenev2013:18175_2419}. 
{\bf Right}: significance map obtained with the {\sc BatImager} software \citep[][]{Segreto10}  in the 20--80~\,keV energy band for 
the SCW 51.  Green circles (24\,$\arcmin$ radius) mark the positions of all sources significantly detected in the field, 
as well as the position of \nosrc\ as reported by \citet{Grebenev2013:18175_2419}.  
In either mosaic, independently built with different software, we do not detect any significant emission from \nosrc.\ 
The bars at the bottom of each mosaic indicate the color codes for the detection significances in units of standard deviations.}  
  \label{sfxt3cand:fig:18175_mosaic}
\end{figure*}

As no significant emission was detected with \sw\,/XRT at the best known position of \nosrc\ 
down to a luminosity that is usually fainter than that of SFXTs in quiescence, 
we reanalysed the \inte\ data in which the source was discovered. 
\inte\ observations are divided into ``science windows'' (SCWs), i.e.\ pointings with typical durations of $\sim$2--3~ks. 
The only source detection is reported in \citet{Grebenev2013:18175_2419},  
where the IBIS/ISGRI data collected in the direction of the X--ray nova 
SWIFT~J174510.8$-$2624 on 2012 September 26 at 14:57 (UT) 
were analysed using the software 
developed at the Space Research Institute of the Russian Academy of Sciences 
\citep[][]{Revnivtsev2004, Krivonos2010:ibis_image_reconstruction}. The detection of the source with the highest significance was 
obtained during the SCW 51 in the satellite revolution 1215. 
The source was also reported to be visible during the first 10 minutes of the SCW 52 in the same revolution, 
albeit with a lower significance. We analysanalyseded these two SCWs by using version 10.2 of the Off-line Scientific Analysis software   
(OSA) distributed by the ISDC \citep{Courvoisier2003}.  

Following \citet{Grebenev2013:18175_2419}, we first extracted the IBIS/ISGRI mosaic of the SCW 51 in the 20--80\,keV energy band and 
searched for the source previously reported at the best determined position of RA(J$2000)$ = $18^{\rm h}\, 17^{\rm m}\,  52\fs8$,  
Dec(J$2000)=-24^{\circ}\,  19^{\prime}\,  45\farcs0$ (the associated uncertainty is 4$\arcmin$). 
The mosaic is shown in Fig.~\ref{sfxt3cand:fig:18175_mosaic} (left). 
We did not detect any significant emission around the position of the source, which lies at an off-axis angle of about 14\,degrees  from
to the center of the instrument FOV. At these large off-axis positions, the calibration of the instrument becomes gradually 
more uncertain and it is difficult to estimate a reliable upper-limit on the X--ray flux. 
By using the {\sc mosaic\_spec} tool, the derived 3$\sigma$ upper limit on any source count-rate at this position 
is of 8.4\,counts\,s$^{-1}$  in the 20--80\,keV energy band (effective exposure time 3281\,s), corresponding to roughy 
40\,mCrab\footnote{The conversion from count-rate to mCrab 
was carried out by using the most recent observations of the Crab (at the time of writing) in spacecraft revolution 1597. 
From these data we measured for the Crab a count-rate of 214.6$\pm$0.3\,counts\,s$^{-1}$ from the IBIS/ISGRI mosaics in the 
20--80~\,keV energy band.} (i.e., 6$\times$10$^{-10}$\,erg\,cm$^{-2}$\,s$^{-1}$). This is 
a factor of 2.5 lower than the flux of the source reported by \citet{Grebenev2013:18175_2419}. 

For completeness, we also extracted the IBIS/ISGRI mosaic of the combined SCW 50, 51, and 52. 
This matches the time interval of the 
light curve showed in Fig.~2 of \citet{Grebenev2013:18175_2419}. 
Also in this case, no significant emission from the position of \nosrc\ is detected. 
The estimated 3\,$\sigma$ upper limit is 15\,mCrab (i.e., 2$\times$10$^{-10}$~erg~cm$^{-2}$~s$^{-1}$) 
in the 20--80\,keV energy band (effective exposure time 9890\,s). 

To further check the source detection we have used an independently developed
software, the {\sc BatImager}, designed to generate sky maps for generic
coded mask detectors and optimised in particular for the processing of
\sw/BAT and \inte/ISGRI data. 
Detailed description and performance of the software when applied on the
BAT survey data is given in \citet[][]{Segreto10}, 
while its imaging performance when applied to the IBIS/ISGRI data are reported 
in \citet[][]{Segreto2008:int,Segreto2010:int}. 
Figure~\ref{sfxt3cand:fig:18175_mosaic} (right) shows the significance map of 
the region around \nosrc\ obtained with the {\sc BatImager} software in the 20--80\,keV energy band 
by using the data in SCW 51. The significantly detected sources in the image are marked with green 
circles, confirming that no source is present at the previously reported position of \nosrc\ \citep{Grebenev2013:18175_2419}. 

Given the above results, we suggest that the source \nosrc\ was erroneously reported. 
This might have occurred due to some mosaic reconstruction problem at the large off-axis 
angle where \nosrc\ should have been located. 
We thus do not discuss this source any further in this paper.

        \section{Discussion and conclusions \label{sfxt3cand:discussion}} 

        \subsection{\xmmj \label{sfxt3cand:disc_2xmm}}

The BAT image trigger on the transient 2XMM J185114.3$-$000004 was a long (64\,s) and  strong (7\,$\sigma$) one. 
The XRT arcsecond position we provide for this source (refinement of 
\citealt{Barthelmy2012:GCN13367}), with an uncertainty of 1\farcs7  at 90\,\% c.l.,   
is consistent with the catalogued position of \xmmj\ 
and is only $0\farcs6$ from 2MASS~J18511447$-$0000036 
($J > 15.6$, $H = 13.23 \pm 0.07$, $K = 11.80 \pm 0.04$). 
This IR source, which we observed with NACO at VLT obtaining 
$J = 15.8 \pm 0.1$ mag and $Ks = 11.7 \pm 0.1$ mag, 
can therefore be safely considered the IR counterpart  of \xmmj\   
\citep[finally removing the need for the cautionary statements on this association 
raised by][]{Bamba2016:2XMM185114}. Even though optical 
spectroscopic data for this source are not yet available and the presence of a supergiant companion 
cannot be firmly established, we show below that all measured properties favour the association of 
\xmmj\ with the SFXT class. 

The BAT light curve of the source (Fig.~\ref{sfxt3cand:fig:2xmm_batxrtlcv}) starts with a bright flare, 
lasting about 100\,s (FWHM, when fit with a Gaussian),
that reached about $\sim10^{-9}$\,erg\,cm$^{-2}$\,s$^{-1}$ (15--50\,keV). 
The overall duration of the outburst is much longer than this (Fig.~\ref{sfxt3cand:fig:2xmm_xrtlcv}), 
as XRT caught several flares in the monitoring observations following the main event. The brightest 
of these flares occurred at $T\sim T_0+1050$\,s and 
reached $\sim8\times10^{-10}$\,erg\,cm$^{-2}$\,s$^{-1}$ (0.2--10\,keV). 
After the first slew, the XRT data show a steep decay lasting until $T+23$\,ks, as typical of the
SFXT population (see Fig.~4, in \citealt[][]{Romano2015:swift10}), 
with a behaviour strongly reminiscent of, e.g., the 2005 August 30 flare of IGR~J16479$-$4514. 
The dynamical range of the initial flare is about three orders of magnitude, but the overall 
soft X--ray dynamical range reaches $\gtrsim$4000 when including the archival \xmm\ data.
We note that, when using a distance of 12\,kpc (see below), the observed luminosities range between
$L \sim 5\times10^{33}$ and $2.6\times 10^{37}$\,erg\,s$^{-1}$. 
This is the typical range of an SFXT source.

The broad band spectrum of \xmmj\ (Fig.~\ref{sfxt3cand:fig:2xmm_batxrtspec}), 
presented here for the first time, 
can be well described by an absorbed cut-off power law model, with 
$\Gamma=0.11_{-0.65}^{+0.68}$ and E$_{\rm f}$=$10_{-3}^{+7}$~keV, 
values which are well within the distribution of parameters measured for HMXB hosting relatively young NSs 
\citep[see ][]{Romano2015:swift10}. 
The measured absorption ($N_{\rm H}=(13\pm4)\times$10$^{22}$~cm$^{-2}$) is much larger than the expected 
Galactic value in the direction of the source \citep[$N_{\rm H}^{\rm Gal}= 1.54\times$10$^{22}$~cm$^{-2}$, ][]{LABS}, 
as expected in SFXTs due to the presence of a dense stellar wind.   
We note (see Fig~\ref{sfxt3cand:fig:2xmmimg}) that there is diffuse emission along the line of sight (LOS) 
to \xmmj\ that may contribute to the reddening. 
This emission is uncorrelated to \xmmj\ and due to the nearby supernova remnant SNR G32.8$-$0.1
as discussed in \citet[][]{Bamba2016:2XMM185114}.

Assuming therefore a blue supergiant nature for the donor star of the \xmmj\ system, 
we can determine the reddening toward
the source by considering its intrinsic NIR color as per 
\citet[][]{Wegner1994}. In the present case, we obtain $J-K$ = 4.1, whereas the 
intrinsic value of this color is $(J-K)_0 \sim$ 0 for early-type 
supergiants, which implies a color excess of $E(J-K) \sim$ 4.1.
Using the Milky Way extinction law of \citet[][]{CCM1989:ABS}, 
this implies a reddening $A_V \approx$ 20 mag, or $A_K \approx$ 2 mag. 
This absorption amount is consistent with the non-detection 
of the object at ultraviolet and optical bands, as found from the UVOT 
data analysis.

This quantity of reddening, when using the formula of \citet[][]{avnh}, 
implies a column density N$_{\rm H} \sim$ 3.6 $\times$ 10$^{22}$ cm$^{-2}$,
which is lower than both that of the Galaxy along the line of sight of
the source (see Sect.~\ref{sfxt3cand:data_2xmm}), 
and that measured in X--rays. This indicates 
that {\it (i)} the object lies within the Galaxy and {\it (ii)} additional X--ray
extinction is present around the accretor, likely produced by the 
accreting material as observed in many other HMXBs and SFXTs.

Concerning point {\it (i)} we can try to estimate the distance of the system  
under the assumption that it hosts an OB supergiant. 
Using $A_V \sim$ 20 mag and the tabulated absolute magnitudes \citep[][]{Lang1992}
and colours \citep[][]{Wegner1994} for this type of stars, we find a distance of 
$\sim$12 kpc. This would 
place the source on the Galactic Plane beyond the Sagittarius-Carina arm 
tangent and close to (or possibly within) the Perseus arm, according 
to the Galaxy map of \citet[][]{Leitch1998}.

        \subsection{\igr \label{sfxt3cand:disc_17407}}

%
%
The nature of the hard X--ray transient \igr\ has been quite controversial since its discovery. 
\citet[][]{Sguera2006} reported three short bright flares reaching $10^{-8}$\,erg\,cm$^{-2}$\,s$^{-1}$ in the 
20--60\,keV band and proposed the association of this object with the SFXT class.  
These authors could not exclude different possibilities, as that of a new ``burst-only''  
source \citep[][]{Cocchi2001,Cornelisse2002}. 
The search for a soft X--ray counterpart for \igr\ led \citet[][]{Heinke2009:chandra} to propose 
CXOU J174042.0$-$280724  
as a likely candidate because of  the observed flaring variability common to both systems. 
Their \chandra\ observations revealed  CXOU J174042.0$-$280724 as a fast transient source, 
varying around a level of 10$^{-13}$--10$^{-12}$\,erg\,cm$^{-2}$\,s$^{-1}$ and thus suggesting the  
presence of an accreting black-hole or NS.  
\citet[][]{Heinke2009:chandra} ruled out the presence of a supergiant companion for any distance larger than  
10\,kpc, and showed that a LMXB or a Be X--ray binary beyond 10\,kpc were more likely possibilities.  
Based on our preliminary position \citep[][]{Romano2011:atel3685}, \citet[][]{Greiss2011:atel3688} 
found a candidate NIR counterpart $0\farcs67$ from the \chandra\ position in archival VVV survey data. 
They find moderate reddening along its LOS and de-reddened optical and NIR magnitudes and colours 
consistent with a late type F dwarf (at a distance of $\sim$3.8 kpc). 
This would thus also disfavour the SFXT hypothesis, unless the F star is a foreground object. 
\citet[][]{Kaur2011:atel3695} 
confirmed the values of the previously reported optical magnitudes and found that the 
candidate counterpart was about one magnitude brighter 4~days after the detected flares observed by \sw.\ 
This is something expected in outbursting LMXBs due to the irradiation of the optical star by the X--rays 
emitted from the compact object.

Thanks to the fact that \igr\ triggered the BAT, we obtained simultaneous soft X--ray 
coverage of the source while it was rapidly decaying, and 
we can now unequivocally establish that the soft X--ray counterpart of \igr\ is indeed 
CXOU J174042.0$-$280724 \citep[see our preliminary results in ][]{Romano2011:atel3685}. 
The \sw/BAT light curve (Fig.~\ref{sfxt3cand:fig:17407_batxrtlcv}a) shows 
at least two bright flares reaching a few 10$^{-9}$\,erg\,cm$^{-2}$\,s$^{-1}$ (15--50\,keV), 
whose profiles are symmetrical and narrow, lasting about 15--20\,s (FWHM, when fitted with a Gaussian shape). 
These values are about a factor of 10 shorter than what measured for typical SFXTs and for \xmmj. 
The second peak (peak2) is also clearly correlated with a soft X--ray flare (Fig.~\ref{sfxt3cand:fig:17407_batxrtlcv}b) 
that reached $\sim2\times10^{-9}$\,erg\,cm$^{-2}$\,s$^{-1}$ (0.5--10\,keV) and that 
lasted at least 10\,s, as the flare is truncated due to the satellite slew, so only a lower limit on its duration is available.
When the source was once again within the XRT FOV, it was already below detection 
(Fig.~\ref{sfxt3cand:fig:17407_xrtlcv}), and all subsequent XRT observations of $\la1$\,ks exposures, 
never revealed the source again, 
yielding individual 3\,$\sigma$ upper limits at the level of 10$^{-12}$\,erg\,cm$^{-2}$\,s$^{-1}$ and a  
combined 3\,$\sigma$ upper limit of $1.1\times10^{-12}$\,erg\,cm$^{-2}$\,s$^{-1}$.  
The serendipitous \xmm\ observations that we analysed show, consistently with what seen in the \chandra\ data 
\citep[][]{Heinke2009:chandra}, a relatively steady flux of a few 10$^{-13}$\,erg\,cm$^{-2}$\,s$^{-1}$ with three 
equally symmetrical and narrow flares reaching 10$^{-12}$\,erg\,cm$^{-2}$\,s$^{-1}$ (0.5--10\,keV) and 
lasting between 190 and 385\,s (FWHM). We note that such flares would go undetected in XRT observations of 
$\sim$1~ks. All the results thus confirmed that this is most likely the truly quiescent behaviour of \igr.  

The source  is therefore characterised by a relatively low, steady flux and infrequent episodes of 
more pronounced activity, with several short bright flares closely spaced in time ($\sim10^3$\,s).  
These flares often reach an X--ray flux that is only a few times higher than the persistent level but more rarely 
can achieve an X--ray dynamic range as high as 4 orders of magnitude.  
As the distance is unknown, the range of observed fluxes would correspond to 
luminosities of $L\sim10^{33}-10^{37}$\,erg\,s$^{-1}$ at 3.8\,kpc (see before), 
or $L\sim10^{34}-10^{38}$\,erg\,s$^{-1}$ at 13\,kpc. 

The \sw\ AT data also provide the first simultaneous broad-band spectroscopy for this transient. 
We obtained satisfactory fits with either an absorbed power law with a high energy roll-over or an absorbed  blackbody. 
The first model provided results that are reminiscent of those measured from SFXTs 
\citep[see, e.g.][for a review of the \sw\ spectra]{Romano2013:Cospar12,Romano2015:swift10}. 
The negative photon index remains, however, puzzling. 
The black-body model provides a quite unlikely high value of the temperature for an accreting object hosted in either a 
HMXB or LMXB. 
In both fits with the power-law with a high energy roll-over and the blackbody, the absorption needed to be fixed to that obtained from the fit 
to the XRT data alone and was consistent with being as low as $N_{\rm H}=(0.84_{-0.84}^{+2.00})\times$10$^{22}$~cm$^{-2}$. 
This is comparable with the expected Galactic 
extinction in the direction of the source \citep[$N_{\rm H}^{\rm Gal}= 0.727\times$10$^{22}$~cm$^{-2}$, ][]{LABS}. 
In principle, this is much lower than the absorption column density expected for an HMXB or an SFXT, thus favouring the LMXB 
hypothesis. However, we note that also the SFXT IGR J08408$-$4503 usually displays an absorption column density 
$\ll$10$^{22}$~cm$^{-2}$ \citep{Bozzo2010:quiesc1739n08408,Sidoli2010:08408}.  
On the other hand, the $N_{\rm H}$ obtained from the XRT data only, as well as from the fit with an absorbed power law with a 
high energy cut-off could still  be consistent with the absorption column density expected for an HMXB or an SFXT. 

The information collected so far on \igr\ is therefore difficult to interpret.  
In the framework of the HMXB/SFXT nature of \igr, pro factors are  
the light curve flaring, which is characterized by at leat two orders of magnitude during the flare 
and the overall dynamic range of about 4 orders of magnitude. The spectral properties are also 
reminiscent of what is usually observed from these systems, even though the broad-band properties of 
these systems are not always a distinctive feature \citep[][]{Romano2014:sfxts_paperX}. 
Against the HMXB/SFXT hypothesis is the low LOS absorption, making it unlikely to hide a supergiant 
behind the F star, even in case the F star is only a foreground source. 
The X--ray flares displayed by \igr\ are also quite short compared to those of the SFXTs, which typically 
last few thousands of seconds. The broad band spectral properties and the optical counterpart could be roughly 
consistent with those of a LMXB but the short flares displayed by \igr\ and the fast variability 
are not commonly observed in LMXBs. The latter typically undergo weeks to months-long outbursts regulated 
by the accretion disk instability \citep{Lasota2001} and might show additional flaring activity on top of 
this \citep[see, e.g.,][for a review]{McClintock2006}. The variability recorded by \sw\ and shown in 
Fig.~\ref{sfxt3cand:fig:17407_batxrtlcv} and \ref{sfxt3cand:fig:17407_xrtlcv} is thus not closely reminiscent 
of that displayed by either blackholes or NS LMXBs in outburst. It is worth noting that the profiles and durations 
of the short flares emitted from \igr\ do not resemble those of type-I X--ray bursts, thus excluding also 
the association between this object and the so-called ``burst only''  sources \citep{Cornelisse2004}. 
Given the relatively high luminosities recorded by \sw\ ($L\sim10^{33}-10^{37}$\,erg\,s$^{-1}$)
for any reasonable estimate of the source distance $\gtrsim$3.8~kpc,
we  consider it an even less likely possibility that \igr\ is a white-dwarf binary \citep[][]{Sazonov2006:XLF}. 
Similarly unlikely is the possibility that \igr\ is a 
a very faint X-ray transient \citep[VFXT,][]{KingWijnands2006:VFXT,Wijnands2006:VFXT}, since 
these objects show a photon index of about 1.5--2.2 in this luminosity range 
\citep[see][]{Degenaar2010:GalCenter,DelSanto2007:VFXT} 
and similar behaviour even when showing hybrid outbursts \citep[faint and bright; see][]{DelSanto2010:sax1753}. 

We conclude that \igr\ is most likely a LMXB hosting an accreting compact object. However, the detailed nature 
of this source remains concealed and spectroscopic 
follow-up of the candidate optical counterpart are encouraged to achieve a more precise classification.

\begin{acknowledgements}
We thank the {\it Swift} team duty scientists and science planners 
for their courteous efficiency, and M.\ Capalbi, M.\ de Pasquale, P.A.\ Evans 
for helpful discussions. 
We also thank the referee for comments that helped improve the paper.
PR, BS, and SC acknowledge contract ASI-INAF I/004/11/0. 
PE acknowledges funding in the framework of the NWO Vidi award A.2320.0076 (PI: N.\ Rea). 
LD acknowledges support by the Bundesministerium f\"ur Wirtschaft und Technologie and
the Deutsches Zentrum f\"ur Luft und Raumfahrt through the grant FKZ 50 OG 1602.
Based on observations made with ESO Telescopes at the Paranal Observatory under programme ID 089.D-0245(A). 
XMM-Newton is an ESA science mission with instruments and contributions directly funded by ESA Member States and NASA.
The XMM-Newton project is supported by the Bundesministerium f\"ur Wirtschaft und Technologie/Deutsches Zentrum 
f\"ur Luft- und Raumfahrt (BMWI/DLR, FKZ 50 OR 1408) and the Max-Planck Society. 
\end{acknowledgements}

\bibliographystyle{aa}

\end{document}